\documentclass[11pt]{article}
\usepackage[T1]{fontenc}

\usepackage{graphicx}
\usepackage{subcaption}
\usepackage[section]{placeins}

\title{Impact of atmospheric effects on the energy reconstruction of air showers observed by the surface detectors of the Pierre Auger Observatory}
\date{}
\begin{document}

\maketitle

\author{
A.~Aab$^{63}$,
P.~Abreu$^{70}$,
M.~Aglietta$^{48,47}$,
I.~Al Samarai$^{29}$,
I.F.M.~Albuquerque$^{16}$,
I.~Allekotte$^{1}$,
A.~Almela$^{8,11}$,
J.~Alvarez Castillo$^{62}$,
J.~Alvarez-Mu\~niz$^{79}$,
G.A.~Anastasi$^{38}$
% ...
\\
\llap{$^{1}$} Centro At\'omico Bariloche and Instituto Balseiro (CNEA-UNCuyo-CONICET), Argentina\\
\llap{$^{2}$} Centro de Investigaciones en L\'aseres y Aplicaciones, CITEDEF and CONICET, Argentina\\
\llap{$^{3}$} Departamento de F\'\i{}sica and Departamento de Ciencias de la Atm\'osfera y los Oc\'eanos, FCEyN, Universidad de Buenos Aires, Argentina\\
\llap{$^{4}$} IFLP, Universidad Nacional de La Plata and CONICET, Argentina\\
\llap{$^{5}$} Instituto de Astronom\'\i{}a y F\'\i{}sica del Espacio (IAFE, CONICET-UBA), Argentina\\
\llap{$^{6}$} Instituto de F\'\i{}sica de Rosario (IFIR) -- CONICET/U.N.R.\ and Facultad de Ciencias Bioqu\'\i{}micas y Farmac\'euticas U.N.R., Argentina\\
\llap{$^{7}$} Instituto de Tecnolog\'\i{}as en Detecci\'on y Astropart\'\i{}culas (CNEA, CONICET, UNSAM) and Universidad Tecnol\'ogica Nacional -- Facultad Regional Mendoza (CONICET/CNEA), Argentina\\
\llap{$^{8}$} Instituto de Tecnolog\'\i{}as en Detecci\'on y Astropart\'\i{}culas (CNEA, CONICET, UNSAM), Centro At\'omico Constituyentes, Comisi\'on Nacional de Energ\'\i{}a At\'omica, Argentina\\
\llap{$^{9}$} Observatorio Pierre Auger, Argentina\\
\llap{$^{10}$} Observatorio Pierre Auger and Comisi\'on Nacional de Energ\'\i{}a At\'omica, Argentina\\
\llap{$^{11}$} Universidad Tecnol\'ogica Nacional -- Facultad Regional Buenos Aires, Argentina\\
\llap{$^{12}$} University of Adelaide, Australia\\
\llap{$^{13}$} Centro Brasileiro de Pesquisas Fisicas (CBPF), Brazil\\
\llap{$^{14}$} Universidade de S\~ao Paulo, Escola de Engenharia de Lorena, Brazil\\
\llap{$^{15}$} Universidade de S\~ao Paulo, Inst.\ de F\'\i{}sica de S\~ao Carlos, S\~ao Carlos, Brazil\\
\llap{$^{16}$} Universidade de S\~ao Paulo, Inst.\ de F\'\i{}sica, S\~ao Paulo, Brazil\\
\llap{$^{17}$} Universidade Estadual de Campinas (UNICAMP), Brazil\\
\llap{$^{18}$} Universidade Estadual de Feira de Santana (UEFS), Brazil\\
\llap{$^{19}$} Universidade Federal de Pelotas, Brazil\\
\llap{$^{20}$} Universidade Federal do ABC (UFABC), Brazil\\
\llap{$^{21}$} Universidade Federal do Paran\'a, Setor Palotina, Brazil\\
\llap{$^{22}$} Universidade Federal do Rio de Janeiro (UFRJ), Instituto de F\'\i{}sica, Brazil\\
\llap{$^{23}$} Universidade Federal Fluminense, Brazil\\
\llap{$^{24}$} Universidad Industrial de Santander, Colombia\\
\llap{$^{25}$} Institute of Physics (FZU) of the Academy of Sciences of the Czech Republic, Czech Republic\\
\llap{$^{26}$} Palacky University, RCPTM, Czech Republic\\
\llap{$^{27}$} University Prague, Institute of Particle and Nuclear Physics, Czech Republic\\
\llap{$^{28}$} Institut de Physique Nucl\'eaire d'Orsay (IPNO), Universit\'e Paris-Sud, Univ.\ Paris/Saclay, CNRS-IN2P3, France, France\\
\llap{$^{29}$} Laboratoire de Physique Nucl\'eaire et de Hautes Energies (LPNHE), Universit\'es Paris 6 et Paris 7, CNRS-IN2P3, France\\
\llap{$^{30}$} Laboratoire de Physique Subatomique et de Cosmologie (LPSC), Universit\'e Grenoble-Alpes, CNRS/IN2P3, France\\
\llap{$^{31}$} Bergische Universit\"at Wuppertal, Department of Physics, Germany\\
\llap{$^{32}$} Karlsruhe Institute of Technology, Institut f\"ur Experimentelle Kernphysik (IEKP), Germany\\
\llap{$^{33}$} Karlsruhe Institute of Technology, Institut f\"ur Kernphysik (IKP), Germany\\
\llap{$^{34}$} Karlsruhe Institute of Technology, Institut f\"ur Prozessdatenverarbeitung und Elektronik (IPE), Germany\\
\llap{$^{35}$} RWTH Aachen University, III.\ Physikalisches Institut A, Germany\\
\llap{$^{36}$} Universit\"at Hamburg, II.\ Institut f\"ur Theoretische Physik, Germany\\
\llap{$^{37}$} Universit\"at Siegen, Fachbereich 7 Physik -- Experimentelle Teilchenphysik, Germany\\
\llap{$^{38}$} Gran Sasso Science Institute (INFN), L'Aquila, Italy\\
\llap{$^{39}$} INAF -- Istituto di Astrofisica Spaziale e Fisica Cosmica di Palermo, Italy\\
\llap{$^{40}$} INFN Laboratori Nazionali del Gran Sasso, Italy\\
\llap{$^{41}$} INFN, Gruppo Collegato dell'Aquila, Italy\\
\llap{$^{42}$} INFN, Sezione di Catania, Italy\\
\llap{$^{43}$} INFN, Sezione di Lecce, Italy\\
\llap{$^{44}$} INFN, Sezione di Milano, Italy\\
\llap{$^{45}$} INFN, Sezione di Napoli, Italy\\
\llap{$^{46}$} INFN, Sezione di Roma ``Tor Vergata``, Italy\\
\llap{$^{47}$} INFN, Sezione di Torino, Italy\\
\llap{$^{48}$} Osservatorio Astrofisico di Torino (INAF), Torino, Italy\\
\llap{$^{49}$} Universit\`a del Salento, Dipartimento di Ingegneria, Italy\\
\llap{$^{50}$} Universit\`a del Salento, Dipartimento di Matematica e Fisica ``E.\ De Giorgi'', Italy\\
\llap{$^{51}$} Universit\`a dell'Aquila, Dipartimento di Scienze Fisiche e Chimiche, Italy\\
\llap{$^{52}$} Universit\`a di Catania, Dipartimento di Fisica e Astronomia, Italy\\
\llap{$^{53}$} Universit\`a di Milano, Dipartimento di Fisica, Italy\\
\llap{$^{54}$} Universit\`a di Napoli ``Federico II``, Dipartimento di Fisica ``Ettore Pancini``, Italy\\
\llap{$^{55}$} Universit\`a di Roma ``Tor Vergata'', Dipartimento di Fisica, Italy\\
\llap{$^{56}$} Universit\`a Torino, Dipartimento di Fisica, Italy\\
\llap{$^{57}$} Benem\'erita Universidad Aut\'onoma de Puebla (BUAP), M\'exico\\
\llap{$^{58}$} Centro de Investigaci\'on y de Estudios Avanzados del IPN (CINVESTAV), M\'exico\\
\llap{$^{59}$} Unidad Profesional Interdisciplinaria en Ingenier\'\i{}a y Tecnolog\'\i{}as Avanzadas del Instituto Polit\'ecnico Nacional (UPIITA-IPN), M\'exico\\
\llap{$^{60}$} Universidad Aut\'onoma de Chiapas, M\'exico\\
\llap{$^{61}$} Universidad Michoacana de San Nicol\'as de Hidalgo, M\'exico\\
\llap{$^{62}$} Universidad Nacional Aut\'onoma de M\'exico, M\'exico\\
\llap{$^{63}$} Institute for Mathematics, Astrophysics and Particle Physics (IMAPP), Radboud Universiteit, Nijmegen, Netherlands\\
\llap{$^{64}$} KVI -- Center for Advanced Radiation Technology, University of Groningen, Netherlands\\
\llap{$^{65}$} Nationaal Instituut voor Kernfysica en Hoge Energie Fysica (NIKHEF), Netherlands\\
\llap{$^{66}$} Stichting Astronomisch Onderzoek in Nederland (ASTRON), Dwingeloo, Netherlands\\
\llap{$^{67}$} Institute of Nuclear Physics PAN, Poland\\
\llap{$^{68}$} University of \L{}\'od\'z, Faculty of Astrophysics, Poland\\
\llap{$^{69}$} University of \L{}\'od\'z, Faculty of High-Energy Astrophysics, Poland\\
\llap{$^{70}$} Laborat\'orio de Instrumenta\c{c}\~ao e F\'\i{}sica Experimental de Part\'\i{}culas -- LIP and Instituto Superior T\'ecnico -- IST, Universidade de Lisboa -- UL, Portugal\\
\llap{$^{71}$} ``Horia Hulubei'' National Institute for Physics and Nuclear Engineering, Romania\\
\llap{$^{72}$} Institute of Space Science, Romania\\
\llap{$^{73}$} University of Bucharest, Physics Department, Romania\\
\llap{$^{74}$} University Politehnica of Bucharest, Romania\\
\llap{$^{75}$} Experimental Particle Physics Department, J.\ Stefan Institute, Slovenia\\
\llap{$^{76}$} Laboratory for Astroparticle Physics, University of Nova Gorica, Slovenia\\
\llap{$^{77}$} Universidad Complutense de Madrid, Spain\\
\llap{$^{78}$} Universidad de Granada and C.A.F.P.E., Spain\\
\llap{$^{79}$} Universidad de Santiago de Compostela, Spain\\
\llap{$^{80}$} Case Western Reserve University, USA\\
\llap{$^{81}$} Colorado School of Mines, USA\\
\llap{$^{82}$} Colorado State University, USA\\
\llap{$^{83}$} Department of Physics and Astronomy, Lehman College, City University of New York, USA\\
\llap{$^{84}$} Fermi National Accelerator Laboratory, USA\\
\llap{$^{85}$} Louisiana State University, USA\\
\llap{$^{86}$} Michigan Technological University, USA\\
\llap{$^{87}$} New York University, USA\\
\llap{$^{88}$} Northeastern University, USA\\
\llap{$^{89}$} Ohio State University, USA\\
\llap{$^{90}$} Pennsylvania State University, USA\\
\llap{$^{91}$} University of Chicago, USA\\
\llap{$^{92}$} University of Hawaii, USA\\
\llap{$^{93}$} University of Nebraska, USA\\
\llap{$^{94}$} University of New Mexico, USA\\
\llap{$^{a}$} School of Physics and Astronomy, University of Leeds, Leeds, United Kingdom\\
\llap{$^{b}$} Max-Planck-Institut f\"ur Radioastronomie, Bonn, Germany\\
\llap{$^{c}$} also at Vrije Universiteit Brussels, Brussels, Belgium\\
\llap{$^{d}$} now at Deutsches Elektronen-Synchrotron (DESY), Zeuthen, Germany\\
\llap{$^{e}$} SUBATECH, \'Ecole des Mines de Nantes, CNRS-IN2P3, Universit\'e de Nantes
}

\begin{abstract}
Atmospheric conditions, such as the pressure
($P$), temperature ($T$) or air density ($\rho \propto P/T$),
affect the development of       extended air showers 
 initiated by energetic cosmic rays. 
 We study  the impact of the atmospheric variations on the reconstruction of air showers with
data from the arrays of surface detectors of the Pierre
 Auger Observatory, considering separately the one with detector spacings of 1500~m  and the one with 750~m spacing.
  We observe modulations in the event rates that are due
to the influence of the air density and pressure variations on the measured signals, from which the energy estimators are obtained. We show how the energy
assignment can be corrected to account for such atmospheric effects.
\end{abstract}

\section{Introduction}
Variation of the atmospheric conditions affect the signals from the extended air shower (EAS)  which can be detected at ground level with arrays of surface detectors, such as the ones of the Pierre Auger Observatory. 
If these effects are not understood and properly accounted for, they can induce systematic effects in the energy reconstruction of the cosmic rays (CRs). 
Consequently, the determination of the CR spectrum and also the search for anisotropies are affected, especially 
 at large angular scales where the daily weather modulations can induce dipolar-like anisotropies in the distribution of arrival directions (although considering time periods of several years, partial cancellations of these modulations are expected). In an earlier investigation 
~\cite{weather1}, we studied  the main effects due to changes in the atmospheric conditions using the data from the surface detector (SD) of the Pierre Auger Observatory up to the end of August 2008. This included a total of about 10$^6$ events of all energies, with a median energy of about 0.6~EeV (where ${\rm EeV}\equiv 10^{18}$~eV). Results were interpreted based on  theoretical models of shower development, validated with simulations of EAS in different atmospheric profiles. Those effects were already taken into account in the analyses of large scale anisotropies performed up to now \cite{LS}.

 In this work  we improve and update  the previous investigation  by including events detected with a four times larger exposure. This  enables us to restrict the dataset to events with energies above 1~EeV, which are less affected by trigger effects, allowing to quantify the atmospheric effects  at the energies  which are used in most of the physics analyses. We also include data from the smaller but denser part of the array of detectors, with 750~m spacing, that was built after the completion of the 1500~m array, for which we consider events with energies above 0.1~EeV. The previous analysis is also improved by including a delay of about two hours in the response of the atmospheric temperature at the relevant heights (of about 500~m to 1~km above ground) with respect  to the changes in the temperature measured at ground level. This inertia of the atmospheric response turns out to be observable with our study of the weather induced modulations of the EAS signals, and is indeed also directly observed in the atmosphere (see the Appendix). Finally, we obtain fits to the zenith angle dependence of the coefficients parameterising the weather induced modulations which are convenient to implement a CR energy reconstruction corrected for the effects of variations in the atmospheric conditions.

\section{The surface detector arrays and the data sets}
The Pierre Auger Observatory is located close to the city of Malarg\"ue in Argentina, at about 1400 m a.s.l.. An essential feature of the Observatory is its hybrid design: cosmic rays above $\simeq 10^{17}$ eV are detected through the observation of the associated air showers with arrays of surface detectors and with fluorescence telescopes overlooking the latter. A detailed description of all components is given in \cite{NIM2015}. Since in the following we will use data from the surface detectors, we briefly recall here their main characteristics. The 1500~m array consists of  1600 water-Cherenkov detectors (WCDs) deployed on a triangular grid, covering a surface of 3000~km$^2$. After the completion of the 1500~m array in 2008, a denser array spaced by 750~m and covering an area of 23.5~km$^2$, nested within the 1500~m array,  was added. Comprising 61~WCDs it extends  the energy range of the 1500~m array, which is fully efficient at energies above 3~EeV, down to lower energies, being fully efficient above 0.3~EeV. The arrays of surface detectors continuously sample the shower particles that reach the ground. The signals registered in the WCDs, due to both the electromagnetic and the muonic components of the EAS, are used to determine the core position and arrival direction of the shower and to determine an estimator of the primary energy. 
For the so-called `vertical' events, i.e., those having zenith angles $\theta<60^\circ$ ($\theta<55^\circ$ for the 750~m array\footnote{We will in the following often write inside parentheses the quantities associated to the 750~m array.}), a fit to the lateral distribution of the signals measured in the 1500~m array (750~m array) is performed to obtain the signal at a reference distance of 1000~m (450~m) from the core. These signals are then converted to the energy estimators $S_{38}$ ($S_{35}$), corresponding to the signals that would have been expected had the shower arrived at a reference zenith angle of 38$^\circ$ (35$^\circ$). The conversion is performed through the method of the constant intensity cut, or CIC in short. This allows one to account for the effects of atmospheric attenuation exploiting the fact that the CR flux is almost isotropic, so that the rate per bin of sin$^2\theta$ above a given true energy should be essentially constant if the detector is fully efficient. In this way one determines a function $f_{\rm CIC}(\theta)$ such that $S_{38}=S/f_{\rm CIC}(\theta)$ (or  $S_{35}=S/f'_{\rm CIC}(\theta)$). Finally, a high-quality subset of hybrid events (i.e., detected simultaneously by the fluorescence and surface detectors) is used to calibrate the SD energy estimators with the energies $E$ measured almost calorimetrically by the fluorescence telescopes. The correlations between the two SD energy estimators and $E$ are well described by a simple power-law function: $E=AS_{38}^B$ ($E=A'S_{35}^{B'}$), in terms of the calibration constants $A$ and $B$ (or $A'$ and $B'$)~\cite{ICRC15}.

\begin{figure}[ht]
  \centering
  	\begin{subfigure}[t]{\textwidth}
  	 \centering
  		\includegraphics[scale=0.7]{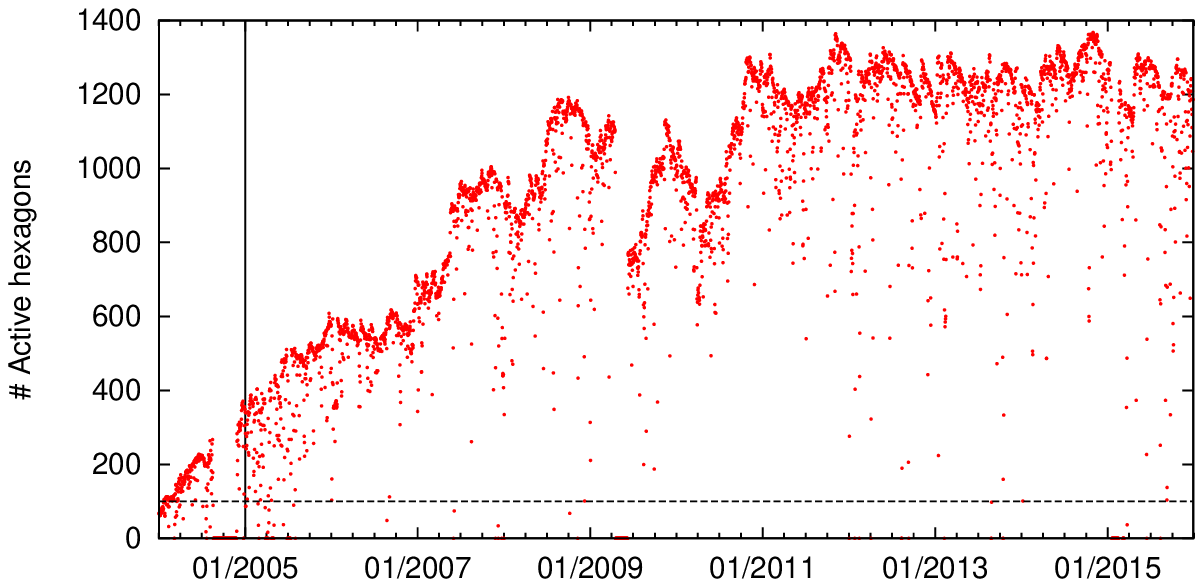}
	\end{subfigure}
	\begin{subfigure}[t]{\textwidth}
	 \centering
    	\includegraphics[scale=0.7]{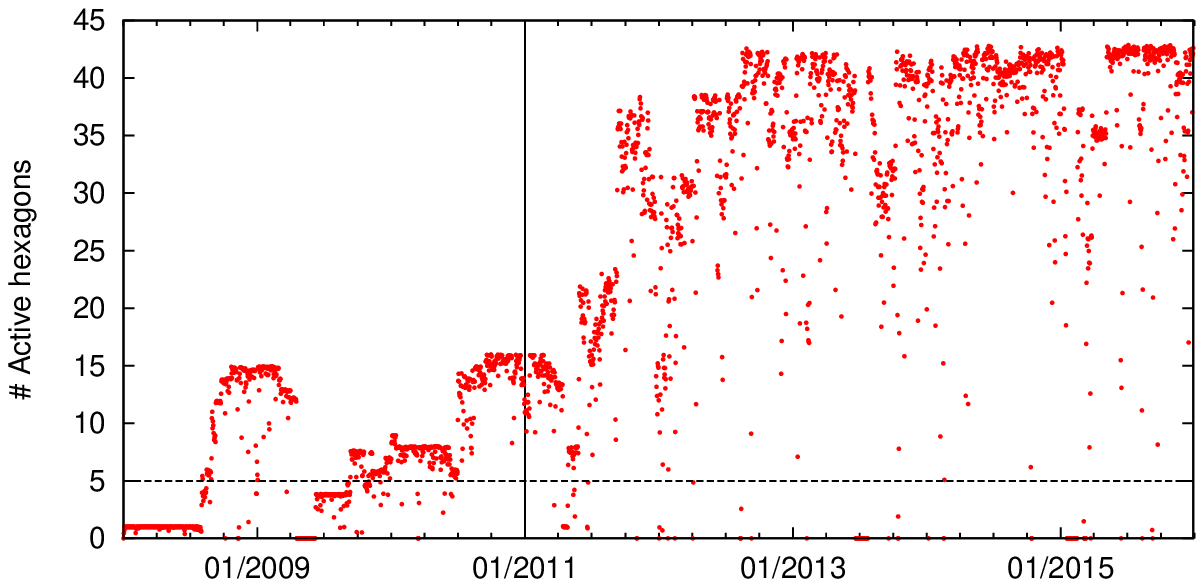}
	\end{subfigure}	
  \caption{Evolution of the number of active hexagons of stations as a function of time for the 1500~m (top) and the 750~m (bottom) arrays. The horizontal dashed lines indicate the value corresponding to about 10\% of the nominal number of hexagons: data periods for which the number of active hexagons is lower are excluded from the analysis.
The vertical lines indicate the starting dates for the present analyses.
}
	\label{fig:hexvst}
\end{figure}

To ensure a good reconstruction of EAS data, only events  which are well contained in the SD array are selected \cite{NIM2010}. This fiducial criterium requires that the detector with the highest signal be enclosed in a hexagon of six active stations. The choice of a fiducial trigger based on active hexagons allows exploiting the regularity of the array to compute the effective area simply as the sum of the areas associated to all active hexagons. Each hexagon contributes a surface of $\sqrt{3}d^2/2$, with $d$ the spacing between the detectors in the equilateral triangular grid. 
The evolution of the number of active hexagons over time, until the end of 2015, is shown in Figure~\ref{fig:hexvst} for the 1500~m array (top panel) and the 750~m array (bottom panel). The 1500~m array started its operation at the beginning of 2004; its size kept growing until its completion in 2008. The 750~m array in turn started to be deployed in 2008 and was completed in 2011. Note that, even after the deployment, the number of hexagons is not constant, due to temporary problems at the detectors (e.g., failures of electronics, power supply, communication system, etc., that are constantly monitored and accounted for in the evaluation of the effective area). In both panels, the vertical lines indicate the starting date for the data used in the present analysis. For the 1500~m array we discard the year 2004, since the array was still quite small and its operation was not very steady. For the 750~m array, in turn, we include only data posterior to January 1 2011, when the operation became more stable and the size became significant with respect to its nominal one.
The horizontal dashed lines in both panels correspond to the minimum number of active hexagons that we consider in this work, i.e., 100 for the 1500~m array and 5 for the 750~m array. As the nominal number is 1380 and 42 for the 1500~m array and the 750~m array, respectively, the chosen values correspond to about 10\% of the nominal ones.

\begin{figure}[ht]
  \centering
  	\begin{subfigure}[t]{\textwidth}
  	 \centering
  		\includegraphics[scale=0.7]{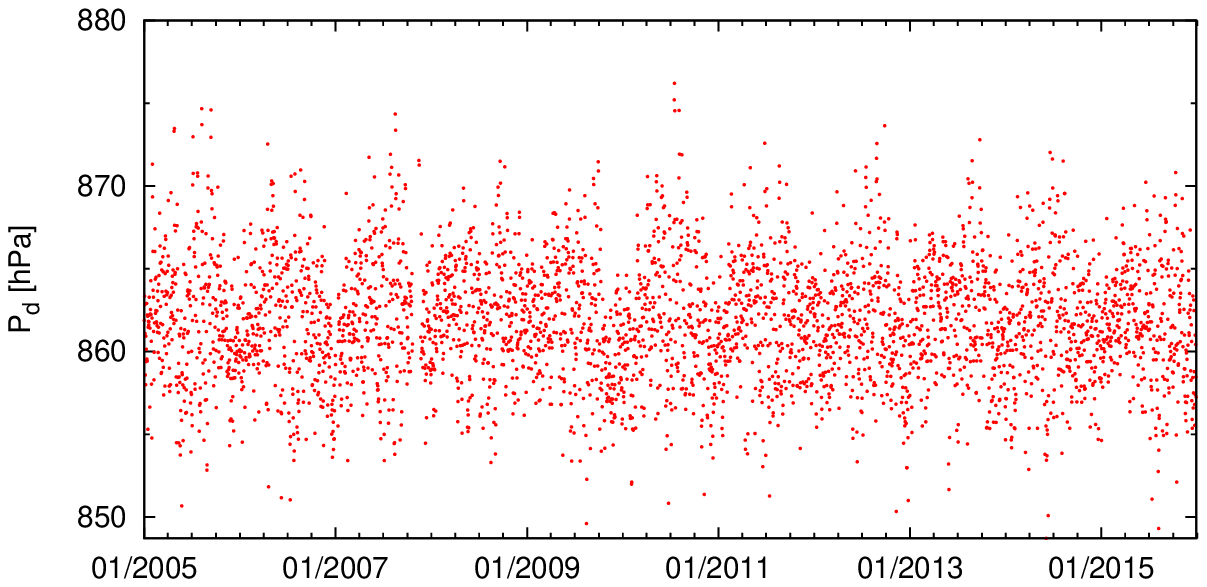}
	\end{subfigure}
	\begin{subfigure}[t]{\textwidth}
	 \centering
    	\includegraphics[scale=0.7]{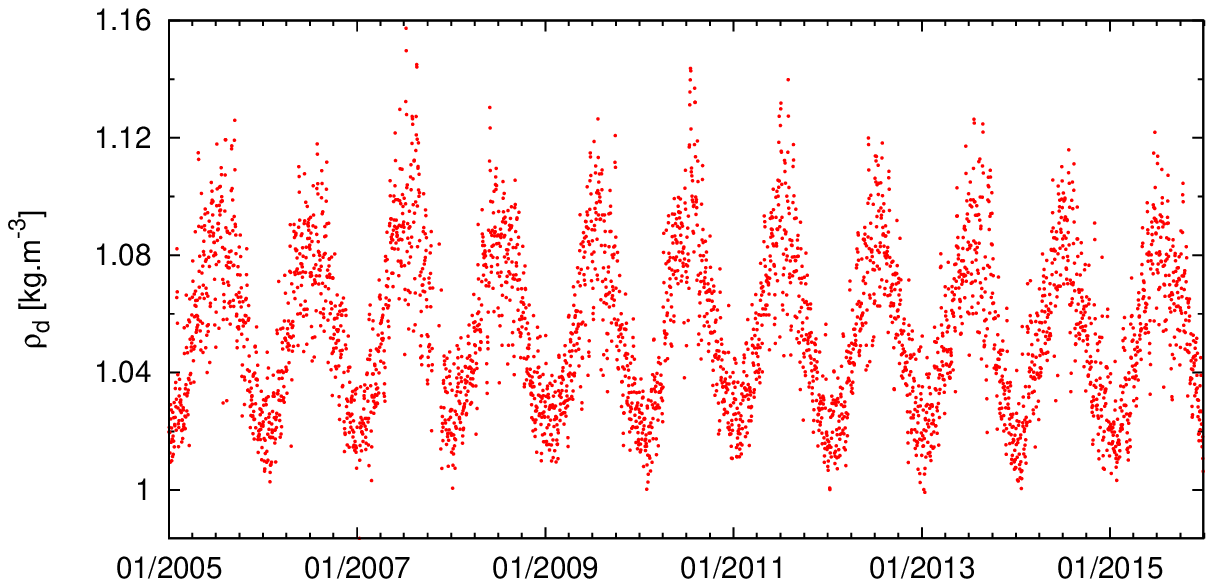}
	\end{subfigure}	
  \caption{Daily averages of $P$ (top) and $\rho$ (bottom).}
	\label{fig:pndvst}
\end{figure}

\begin{figure}[ht]
  \centering
  	\begin{subfigure}[t]{\textwidth}
  	 \centering
  		\includegraphics[scale=0.7]{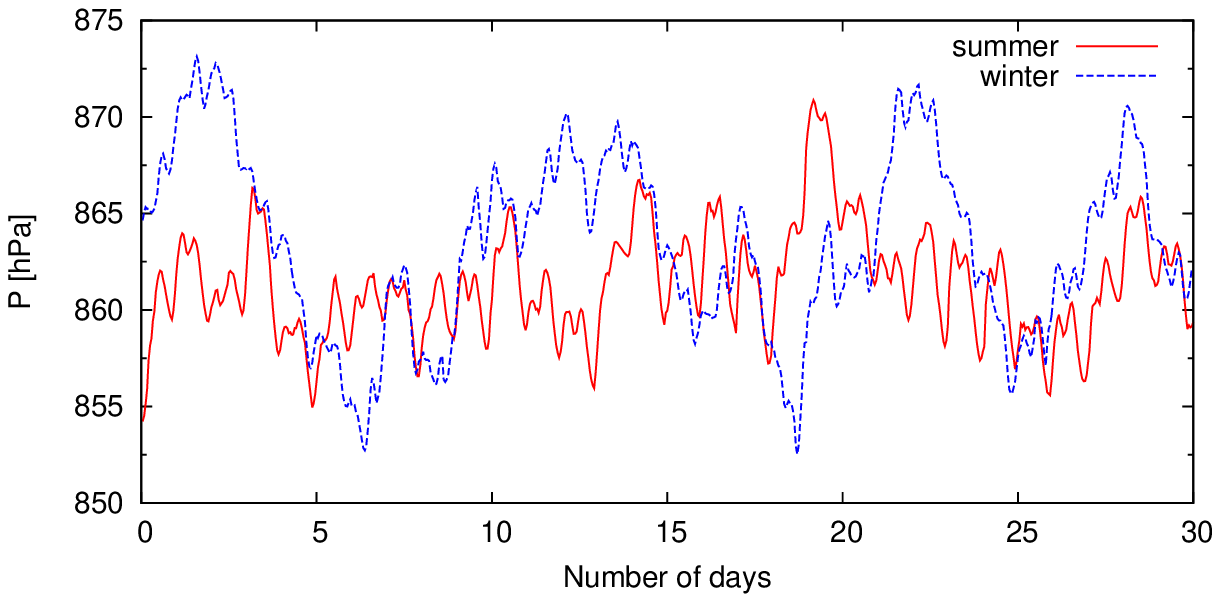}
	\end{subfigure}
	\begin{subfigure}[t]{\textwidth}
	 \centering
    	\includegraphics[scale=0.7]{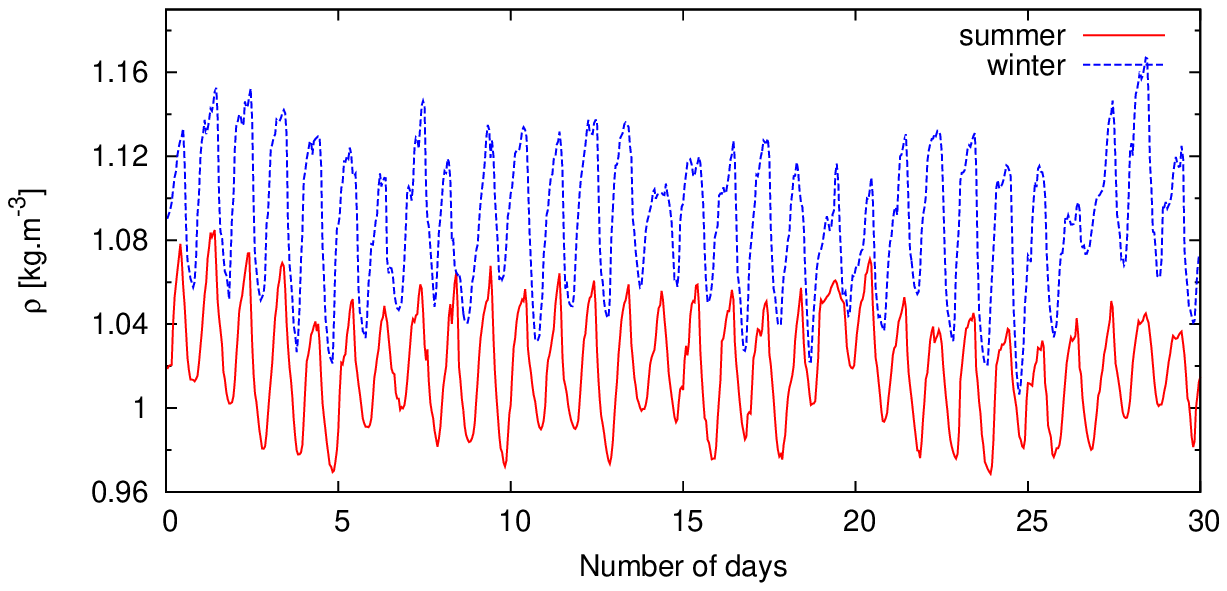}
	\end{subfigure}	
  \caption{Measurements at the CLF weather station of $P$ (top) and $\rho$ (bottom) every 5~min  over 30 days in a summer month, January 2015 (solid) and a winter month, July 2014 (dashed).}
	\label{fig:pndvst30d}
\end{figure}

Several weather stations are in operation at the Auger Observatory to monitor the atmospheric conditions, four at the different sites of the fluorescence telescope buildings and one near the laser facility close to the centre of the 1500~m array \cite{NIM2015}. For the present analysis, we use a database consisting of the temperature and pressure measured at the Central Laser Facility (CLF) weather station, available every 5~minutes for most of the time. When gaps in the data between 10 minutes up to three hours are present, the data are just interpolated from the values at the endpoints of those empty intervals. For the longer periods when this station is not operational, we adopt data from the other stations\footnote{Since weather stations lie at different heights (the stations used are CLF at 1401~m (from which 92.8\% of the atmospheric data is retrieved), Los Leones at 1420~m (3.8\% of the data), Los Morados at 1423~m (3.0 \% of the data) and Loma Amarilla at 1483~m (0.4\% of the data)), the pressure measurements are corrected to the reference height of CLF. Also note that temperature differences of up to $\pm 3$ degrees can appear between stations due, e.g., to different cloud coverage, but their average effect is negligible .}, if available, or otherwise discard the period for the present analysis. 

As discussed in the next section, the most convenient way to study the atmospheric effects on EAS is by considering as variables the pressure, $P$, and the air density, $\rho$. The latter is approximately determined from the values of $P$ and $T$ through the dry air relation $\rho\simeq  0.3484 P/(T+273.16)$~kg\,m$^{-3}$, with $P$ in hPa and $T$ in $^\circ$C, with the humidity having only a subdominant effect in the density near ground. In Figure~\ref{fig:pndvst} we show the values at the CLF location of the daily-averaged pressure (top panel) and density (bottom panel) during the time periods considered in this work, being the uncertainties in these averaged quantities negligible. In Figure~\ref{fig:pndvst30d} we plot with red lines the actual pressure and density measured every 5 min in one month during austral summer (January 2015 as an example) and with blue dashed lines in a winter month (July 2014).  
The seasonal modulation in the density is quite apparent, being anti-correlated with the temperature changes, so that it reaches a maximum in winter and a minimum in summer. Overall, the daily-averaged density changes by about $\pm6$\% during the year, while the changes within a day can be $\pm 3$\% (corresponding to temperature changes of $\pm 8^\circ$C). Regarding the pressure, the seasonal modulation is less noticeable but its variations are more pronounced in winter times than during the summer.
On a daily basis, the pressure at the Malarg\"ue site is found to have generally a minimum in the afternoon, and a shallower local minimum in the early morning, and it  is moreover significantly modulated by the movement of cold fronts and storm activity  at the site, which are more frequent during winter times. 

\begin{table}[ht]
\centering
\begin{tabular}{l c c}
\hline
\hline
 & 1500~m array & 750~m array\\
\hline
Time period & 1/1/2005 - 31/12/2015 & 1/1/2011 - 31/12/2015\\
Exposure [km$^2$ sr yr] & $46,438$ & $159$\\
Energy Threshold [EeV] & 1.0 & 0.1\\
Number of events & $1,146,481$ & $570,123$\\
Median energy  [EeV]& 1.5 & 0.15 \\
\hline
\hline	
\end{tabular}
\caption{Characteristics of the data sets used in the analyses.}
\label{tab:datasets}
\end{table}

To conclude this section, we summarize in Table~\ref{tab:datasets} the characteristics of the two data sets used in this work, including the time periods considered and the relative exposures. The number of events above the given energy threshold is that obtained after applying to data the selection cuts described above, based on fiducial criteria, minimum array size and data periods with measured weather data.

\section{Determination of atmospheric coefficients from the modulations of the event rate}

In ref.~\cite{weather1} we have described in detail the effects of atmospheric variations on EAS as observed at the Pierre Auger Observatory. We remind here the aspects relevant to the analysis performed in the present work. The two atmospheric variables that mostly affect the shower development are the pressure and the air density. On the one hand, the pressure, which essentially measures the density of the vertical column of air, determines the ``age'' of the shower when it reaches the ground. Consequently, its variations affect the observed signals because of the attenuation of the longitudinal shower profile as the slant depth increases beyond that of the shower maximum, reducing the signal as the shower gets older. On the other hand, variations of the air density affect, via the Moli\`ere radius, the lateral spread of shower particles due to multiple Coulomb scattering of the electromagnetic component of the shower. This  affects the measured signals, which become smaller as the density increases. 
An additional issue arises because the electromagnetic signal at the reference distance from the core is produced typically by the shower development over the last two cascade units (i.e., radiation lengths in air) before hitting the ground, corresponding to heights of 500~m to 1~km above ground level. 
This has two main implications. First, although the density (or temperature) variations at those heights are correlated with the ones determined at ground level, the daily amplitude of the temperature variations are smaller above ground level than at ground level, typically by a factor 1/2 to 1/3 at the relevant heights considered. Second, there is a delay of about two hours in the response of the atmosphere to the temperature variations  produced by  the heating or cooling of the ground.
An effective way to account for these effects is to split the modulation due to the density variations into two terms: the first term is proportional to the variations of the daily averaged density, which has similar variations at ground level and at heights of 1~km above it; the second term is  proportional to the deviation of the density at a given time from the daily average, which is  smaller at the relevant heights than at ground level. 
This daily modulation should also be retarded by about two hours in order to account for the inertia of the atmosphere to the  changes in temperature determined at ground level.

For each reconstructed EAS event, the signals $S$ at the reference distance from the shower axis of 1000~m (450~m), obtained from the fit to the  lateral distribution of the signals measured in the individual WCDs, are expected to be modulated according to

\begin{equation}
S=S_0[1+\alpha_P(P-P_0)+\alpha_\rho(\rho_d-\rho_0)+\beta_\rho(\tilde\rho-\rho_d)],
\label{svspr}
\end{equation}
where $P_0=862$ hPa and $\rho_0=1.06~\mathrm{kg~m}^{-3}$ are reference values for $P$ and $\rho$ corresponding to averages at the site of the Pierre Auger Observatory during the years considered here. $S_0$ is the value of the signal that would have been obtained at those reference atmospheric  conditions, $\rho_d$ is the daily average of the density (within $\pm 12$~h of the event) while $P$ and $\rho$ are the actual pressure and density determined at ground level at the time of the event, $\tilde\rho$ being the density that was determined two hours before. The coefficients $\alpha_P$, $\alpha_\rho$ and $\beta_\rho$ parameterize the assumed linear dependence of the signal modulation.

The  dependence of the signal on the atmospheric conditions leads to a modulation of the rate $R$ of recorded events above a given signal $S_{\rm min}$, per unit time, area and zenith angle, which can be written as

\begin{equation}\label{eq::evrate1}
\frac{{\rm d}R}{{\rm d}\theta}=2\pi\sin \theta \cos \theta\int_{S_{\rm min}}^\infty {\rm d}S\,P_{\rm tr}(S,\theta)\frac{{\rm d}\Phi_{CR}}{{\rm d}E_t} \frac{{\rm d}E_t}{{\rm d}S},
\end{equation}
where $P_{\rm tr}$ accounts for a possible non-saturated trigger efficiency for energies $E<3$~EeV ($E<0.3$~EeV), assumed to depend, for every given zenith angle, just on the signal $S$ at the reference distance. The  differential flux of cosmic rays per unit solid angle is assumed to follow a power law $E_t^{-\gamma}$, with $E_t$ the true energy of the CRs, and the spectral index will be taken here as $\gamma=3.29$, as determined by the Auger Observatory \cite{ICRC15} at energies below $\sim 5$~EeV (i.e., below the ankle feature above which the spectrum hardens). This spectral index is the one relevant for this study since in the following we will consider threshold energies of 1~EeV (or 0.1~EeV for 750~m array). Using now the energy calibration relation with $E_t\simeq AS_0^B$ in terms of the signal $S_0$ at the reference atmospheric conditions (as will be further discussed in Section~4) we obtain from Eq.~(\ref{svspr}),  expanding to first order in the weather corrections, that 

\begin{equation}\label{eq::evrate2}
\frac{{\rm d}R}{{\rm d}\sin^2\theta}\propto [1+a_P(P-P_0)+a_\rho(\rho_d-\rho_0)+b_\rho(\tilde\rho-\rho_d)]\int_{S_{\rm min}}^\infty  {\rm d}S\,P_{\rm tr}(S,\theta)S^{-B\gamma+B-1}.
\label{rate.eq}
\end{equation}

For the assumed CR energy spectrum with the shape of a  power-law,  the relation between the rate coefficients and the signal ones is  $a_{P,\rho}=B(\gamma-1)\alpha_{P,\rho}$ and similarly $b_{\rho}=B(\gamma-1)\beta_{\rho}$.  The coefficient $B$ is derived from the energy calibration ~\cite{ICRC15} and is equal to $B=1.023\pm 0.006$ for the 1500~m array  ($B'=1.013\pm 0.013$ for the 750~m array). One has then that $a_{P,\rho}\simeq 2.3\alpha_{P,\rho}$ and similarly $b_{\rho}\simeq 2.3\beta_{\rho}$.

To determine the atmospheric coefficients we compute the rate by counting the  events in one hour bins and normalizing to the corresponding area of the array at that time, which is calculated  from the total number of hexagons of active neighboring detectors, which is known at every second. 

We finally use the expression given by Eq.~(\ref{eq::evrate2}) to fit the measured rate of events. Assuming that the number of events $n_i$ observed in each hour bin $i$ follows a Poisson distribution of average $\mu_i$, a maximum likelihood fit is performed to estimate the coefficients $a_P$, $a_\rho$ and $b_\rho $.
The likelihood function is $L=\prod {\mu_i^{n_i}}e^{-\mu_i}/{n_i!}$. The expected number of events in bin 
$i$ is given by \[\mu_i=R_0\times A_i\times C_i,\] where $R_0$ is the average rate that would have been observed if the atmospheric parameters were always the reference ones, i.e., $R_0={\sum n_i}/{\sum A_i C_i}$, with $A_i$ the sensitive area in the \emph{ith} time bin and \[C_i=1+a_P(P_i-P_0)+a_\rho(\rho_{d_i}-\rho_0)+b_\rho(\tilde{\rho}_i-\rho_{d_i}),\]
where $\tilde{\rho}_i=\rho_{i-2}$ means that the density used in the fitting procedure is the one measured two hours before.

\subsection{Results for the 1500~m array}

For the 1500~m array a first analysis was performed in ref.~\cite{weather1} using data from January 1 2005 up to August 31 2008, with no energy selection other than that imposed by the trigger,  corresponding to about $10^6$ events with median energy of about 0.6~EeV. We here extend this dataset up to the end of 2015. Given the much larger statistics available it becomes possible to determine the atmospheric coefficients using a higher threshold of 1~EeV, with a median of about 1.5~EeV, so that the rates are less affected by trigger effects at energies below full trigger efficiency. This also leads to coefficients determined in an energy range  which is closer to the energies at which the physics analyses are performed. One has to keep in mind that the coefficients may depend on the energy, due e.g., to the logarithmic energy dependence of the depth of shower maximum, the possible energy dependence of the slope of the lateral distributions of signals measured at ground or of the electromagnetic fraction of the shower, or even due to changes in the predominant CR composition at different energies. Anyhow, this energy dependence is not expected to be large, so that even at the highest energies the correction performed using the coefficients determined close to the EeV should already account for most of the effects.

\begin{figure}[ht]
  \centering
  	\begin{subfigure}[t]{\textwidth}
  	  \centering
  		\includegraphics[scale=0.65]{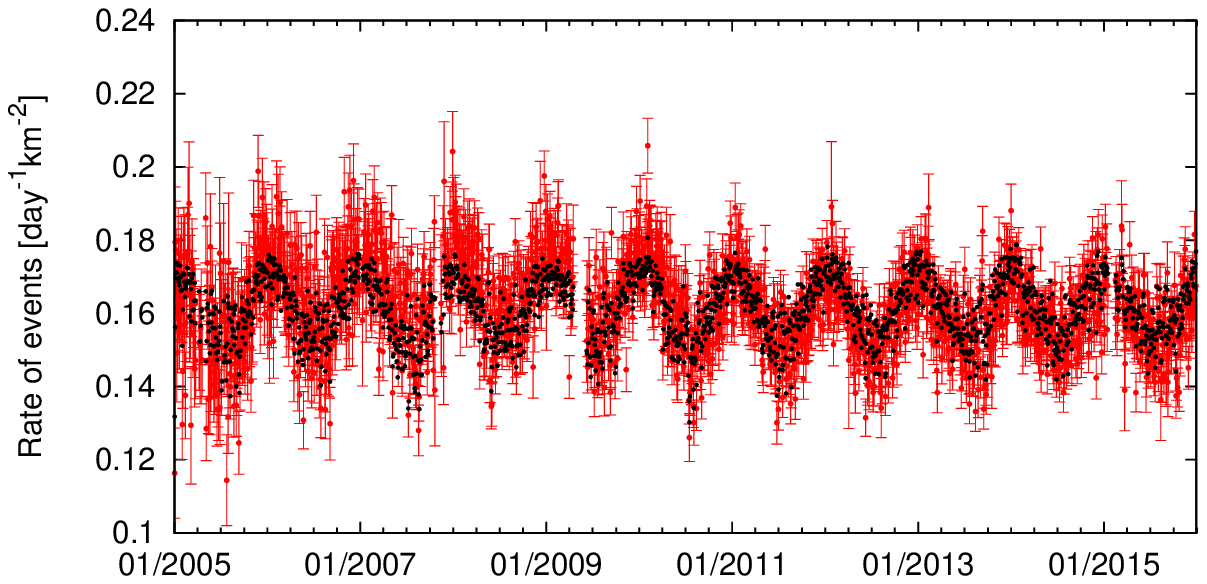}
	\end{subfigure}
	\begin{subfigure}[t]{\textwidth}
	  \centering
    	\includegraphics[scale=0.65]{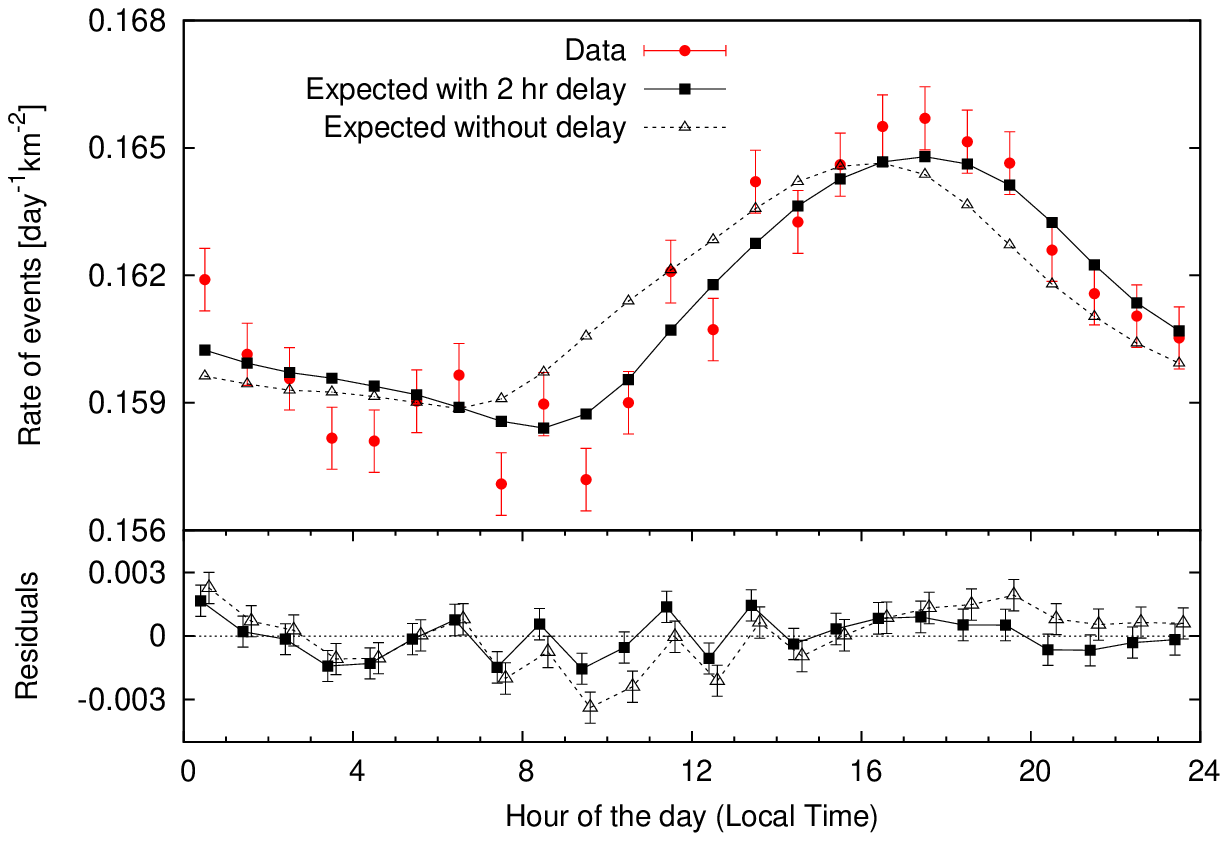}
	\end{subfigure}	
  \caption{Top panel: Rate of events per day, with $E>1$~EeV, for the 1500~m array. The experimental points are shown in red while the black points represent the expectations from the fit. Bottom panel: Average hourly measured rates and expectations from the  fit including the two hour delay (squares) or with the actual density (triangles), with their residuals. }
	\label{fig:ratema}
\end{figure}

 We first perform a fit to the atmospheric coefficients including all zenith angles,  $\theta<60^\circ$,
 by using Eq.~(\ref{rate.eq}) to fit the rate of events with $E>1$~EeV in one hour bins. In this way we obtain the following (zenith angle averaged) coefficients:
\begin{eqnarray}\label{eq::fitpMA}
a_P &=& (-3.2 \pm 0.3)\times 10^{-3}~\mathrm{hPa}^{-1}\nonumber\\
a_\rho &=& (-1.72 \pm 0.04)~\mathrm{kg}^{-1}\mathrm{m}^3\\
b_\rho &=& (-0.53 \pm 0.04)~\mathrm{kg}^{-1}\mathrm{m}^3.\nonumber
\label{aver1500}
\end{eqnarray}
All errors are the statistical ones associated to the fit.
The reduced $\chi^2$ obtained is $\chi^2/dof=1.013$ (for 88,126 degrees of freedom), where $\chi^2=\sum_i (n_i-\mu_i)^2/\mu_i$.

\begin{figure}[t]
  	\begin{subfigure}[t]{0.32\textwidth}
  		\includegraphics[scale=0.2]{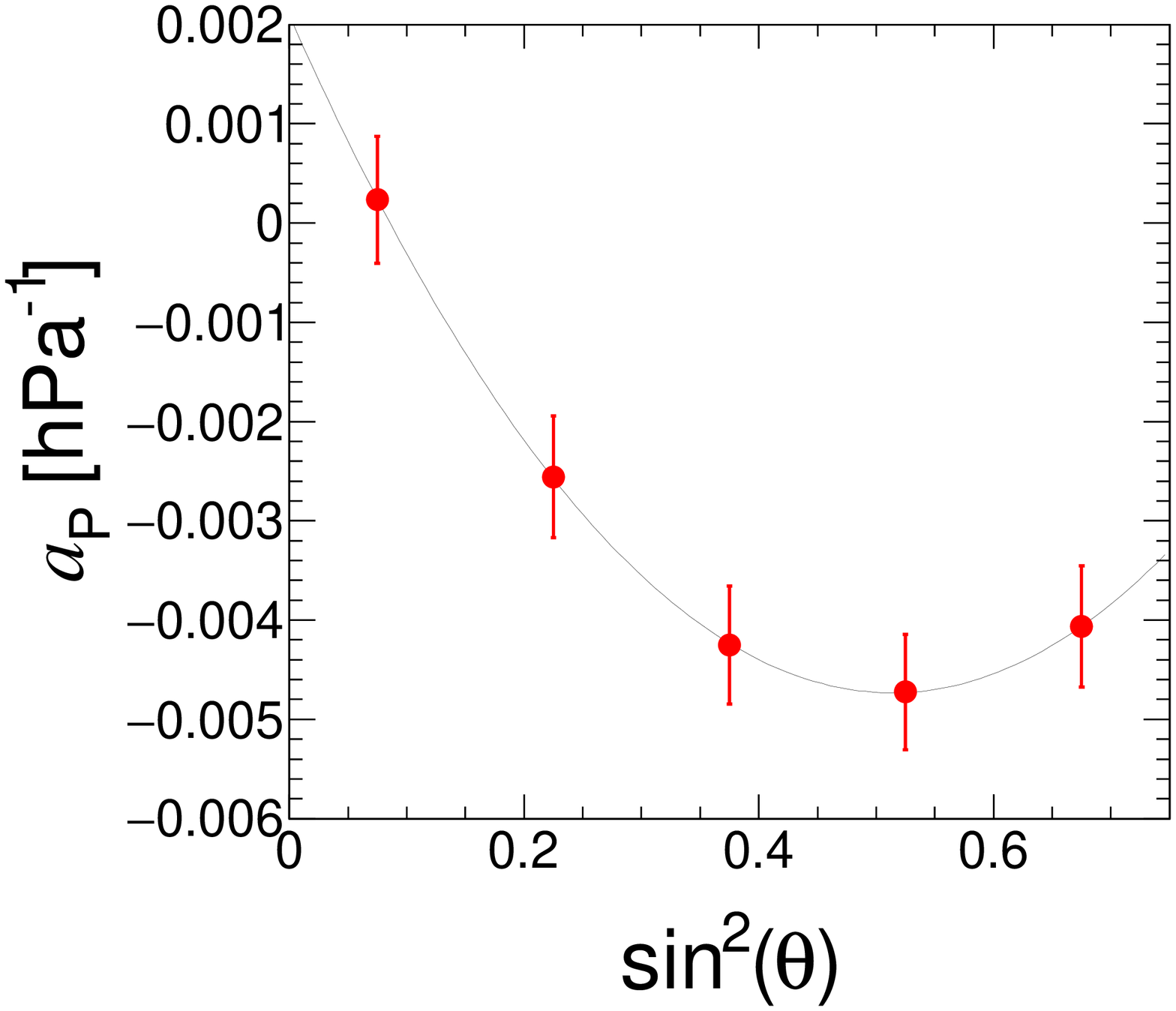}
	\end{subfigure}
	\begin{subfigure}[t]{0.32\textwidth}
    	\includegraphics[scale=0.2]{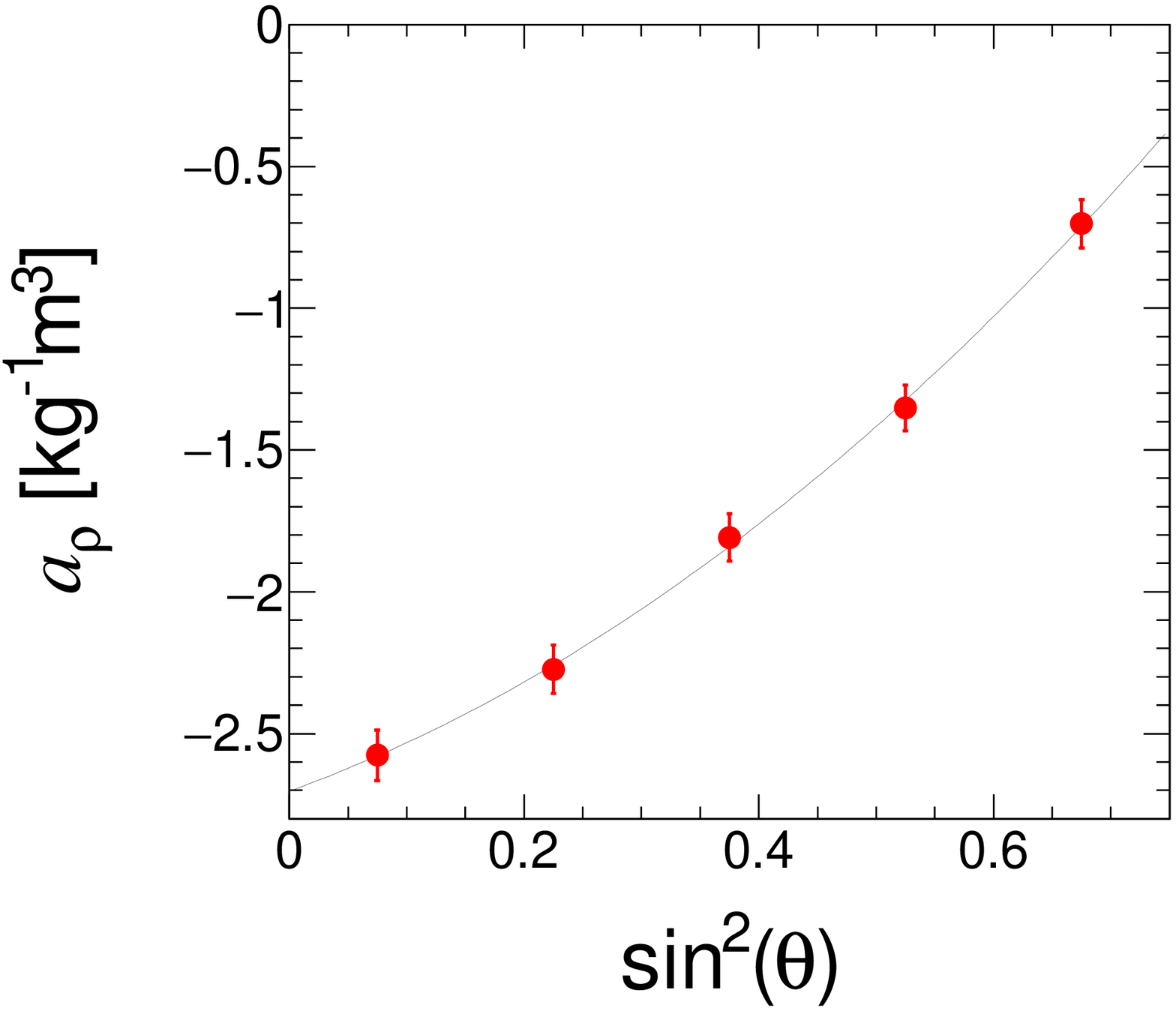}
	\end{subfigure}	
	\begin{subfigure}[t]{0.32\textwidth}
    	\includegraphics[scale=0.2]{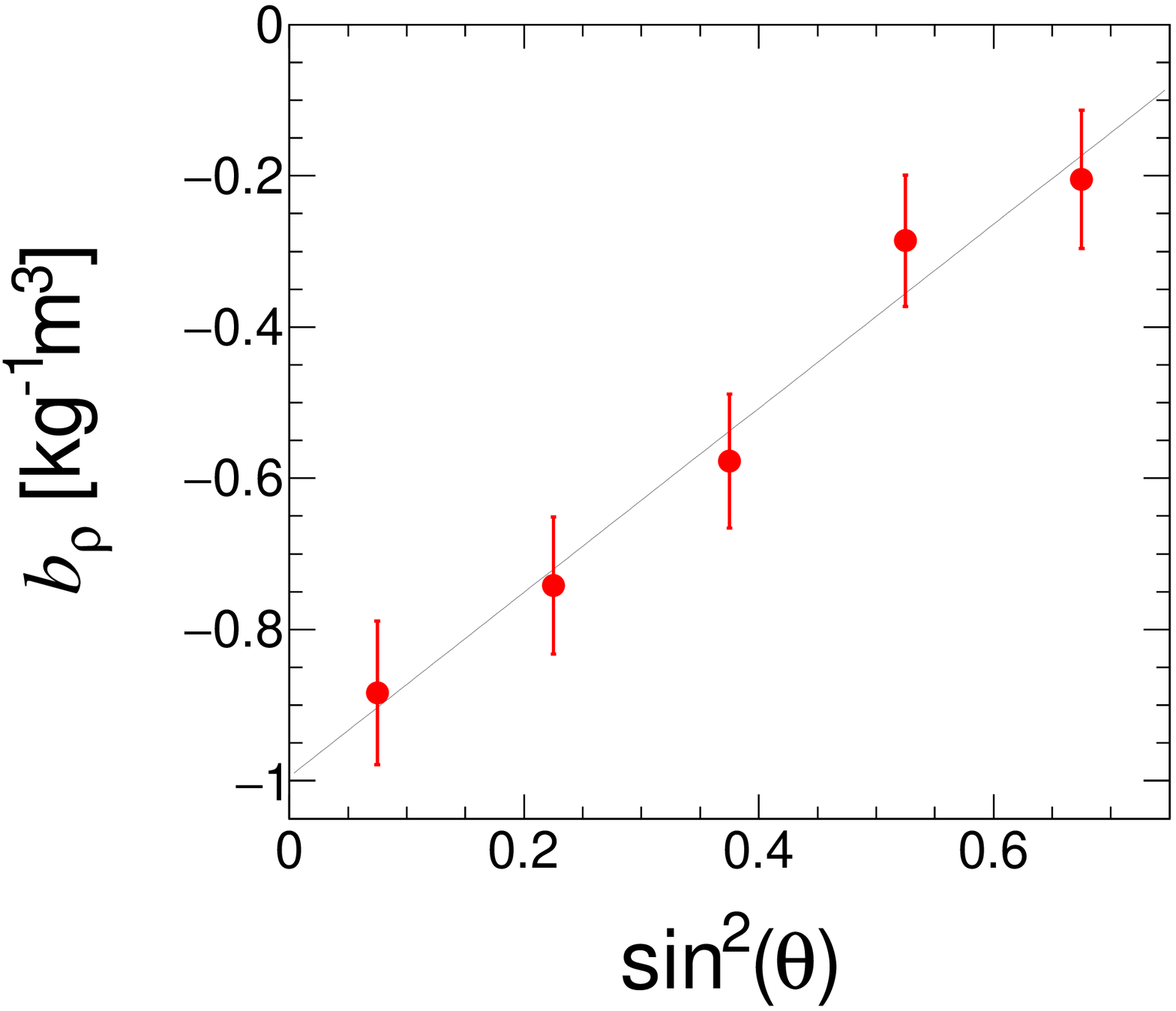}
	\end{subfigure}	
  \caption{Atmospheric coefficients $a_P$ (left), $a_\rho$ (middle) and $b_\rho$ (right) as a function of $\sin^2\theta$,   for the 1500~m array. 
  The line is the fit obtained using a  quadratic polynomial to describe the dependence on sin$^2\theta$.}
	\label{fig:wcvssin2ma}
\end{figure}

The behavior of the daily averaged rate (i.e., in 24~h bins) over the time period considered is shown in Figure~\ref{fig:ratema}, top panel (red points and corresponding error bars).
The black points represent the rates expected according to the coefficients derived from the fit. The agreement between measured and expected rates is evident, as expected from the goodness of the fit. The bottom panel displays the average rates as a function of the hour of the day (the local time at Malarg\"ue corresponds to UTC--3, having no daylight saving changes over the year). Here red points with error bars are the data, squares represent the expectation from the fitted coefficients while triangles would be the expectations if one were to use in the expression in Eq.~(\ref{svspr}) the actual air density $\rho$  rather than the density $\tilde\rho$ measured two hours before. 
A significant improvement is obtained when including the delay in the fit,  with the reduced $\chi^2$ changing from 4.2 without delay  down to 1.9 with delay (for 21 degrees of freedom). The residuals in both cases are shown in the lower insert. It is apparent that the improvements obtained with the delay are most noticeable at the times of the day at which the temporal variation in the temperature is maximal. To guide the eye we joined with lines the points corresponding to the expectations from the model.

To study the dependence of the atmospheric coefficients on the zenith angle we divide the data set into five bins of equal width in $\sin^2\theta$, so that the number of events per bin is similar, and fit the coefficients in each subset\footnote{Note that a binning in sin$^2\theta$ is also used to fit the CIC functional dependence. In ref.~\cite{weather1} we used instead bins in sec$\theta$, but this leads to significantly smaller number of events, and hence larger statistical errors, for large values of sec$\theta$.}.   
The results are shown in Figure~\ref{fig:wcvssin2ma}, together with a fit to the zenith angle dependence of the coefficients using quadratic polynomials

\begin{table}[t]
\centering
\begin{tabular}{ c c c c }
\hline\hline
 & $c_0$ & $c_1$ & $c_2$\\
\hline\\[-1.5ex]

$a_P~[{\rm hPa}^{-1}]$ & $(2.1 \pm 0.9)\times 10^{-3}$ & $(-2.6 \pm 0.6)\times 10^{-2}$ & $(2.6 \pm 0.7)\times 10^{-2}$\\
$a_\rho~[{\rm kg}^{-1}{\rm m}^3]$ & $-2.7 \pm 0.1$ & $1.5 \pm 0.8$ & $2.2 \pm 1.0$\\
$b_\rho~[{\rm kg}^{-1}{\rm m}^3]$ & $-1.0 \pm 0.1$ & $1.2 \pm 0.8$ & $0.0 \pm 1.1$\\
\hline\hline
\end{tabular}
\caption{Parameters of the fits to the zenith angle dependence of the atmospheric coefficients for the 1500~m array, using the quadratic polynomials in Eq.~(\ref{eq::fitwc}).}
\label{tab:wcvssinfitma}
\end{table}

\begin{equation}\label{eq::fitwc}
f(x)=c_0+c_1 x+c_2x^2\ \ \ ,\ \ \ {\rm where } \ x=\sin^2\theta.
\end{equation}
Here $f$ might be $a_P$, $a_\rho$ or $b_\rho$. The resulting values of the coefficients are summarized in Table~\ref{tab:wcvssinfitma}.  The main features of the results obtained can be understood by noting that the pressure coefficient is negative because an increase in
atmospheric pressure corresponds to an increase in the vertical column of matter traversed by the
shower, hence the same EAS will be observed at a later stage of development as the pressure
increases. The signals are actually the superposition of the electromagnetic and the muonic components, and while the latter changes little with increasing depth, the electromagnetic one gets exponentially suppressed beyond the shower maximum\footnote{Moreover, the maximum of the electromagnetic component at 1000~m from the core is deeper by about 100~g\,cm$^{-2}$ than the maximum at the core which is measured by the fluorescence detectors \cite{weather1}.}. For the energies considered here the shower maximum at 1000~m from the core is close to ground level for vertical showers, which explains the small value of $a_P$ for $\theta\simeq 0^\circ$. For increasing zenith angles, the shower reaches ground when the longitudinal development is already getting suppressed. Hence, the energy estimator becomes smaller when the pressure increases. This effect gets more pronounced as the zenith angle increases until for zenith angles approaching 60$^\circ$ the electromagnetic component starts to become subdominant and hence the pressure coefficient starts to become smaller in absolute value.  Regarding the density coefficients, $a_\rho$ and $b_\rho$, they are also negative since a larger air density reduces the lateral spread of the shower. For large zenith angles they become smaller due to the suppression of the electromagnetic fraction of the signal. One can also appreciate that the coefficient $b_\rho$ is smaller than $a_\rho$ by a factor of about one third, due to the proportionally smaller amplitude of daily temperature variations at the relevant heights with respect to those at ground level.

\subsection{Results for the 750~m array}
\ 

Considering now the 750~m array  and
 using Eq.~(\ref{rate.eq}), we fit the rate of events with $E>0.1$~EeV and $\theta<55^\circ$, computed in one hour bins, to obtain the coefficients averaged over zenith angles:
\begin{eqnarray}\label{eq::fitpIN}
a_P &=& (-4.9 \pm 0.4)\times 10^{-3}~\mathrm{hPa}^{-1}\nonumber\\
a_\rho &=& (-1.07 \pm 0.06)~\mathrm{kg}^{-1}\mathrm{m}^3\\
b_\rho &=& (-0.37 \pm 0.06)~\mathrm{kg}^{-1}\mathrm{m}^3\nonumber.
\label{aver750}
\end{eqnarray}
The reduced $\chi^2$ of the fit is $0.998$ (for 39,258 degrees of freedom). 

\begin{figure}[ht]
  \centering
  	\begin{subfigure}[t]{\textwidth}
  	  \centering
  		\includegraphics[scale=0.65]{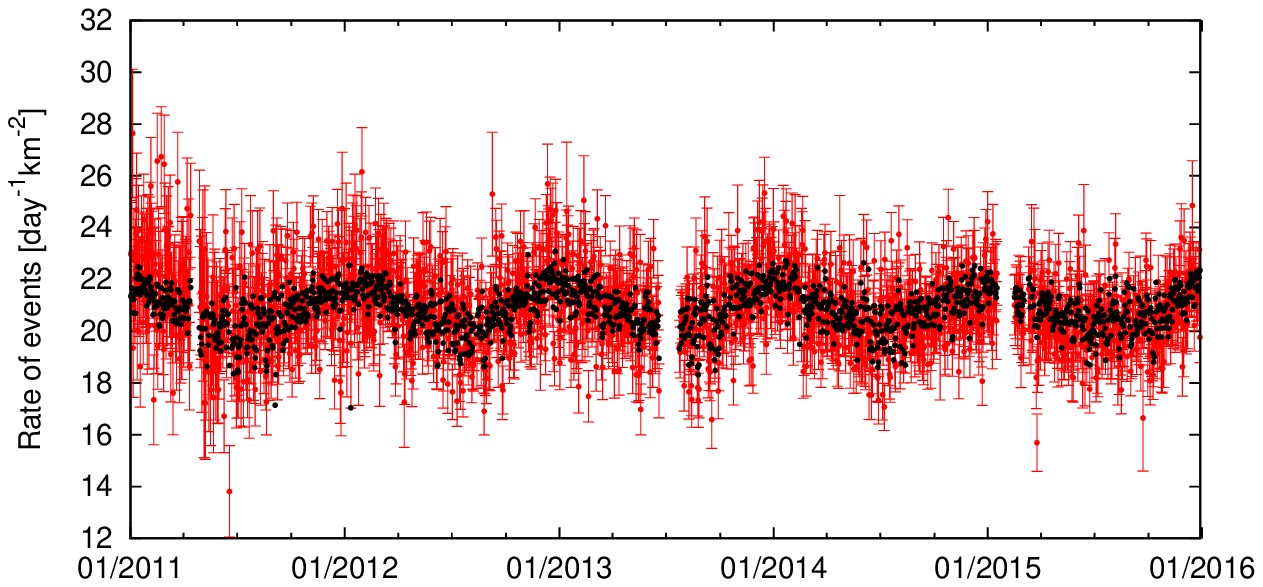}
	\end{subfigure}
	\begin{subfigure}[t]{\textwidth}
	  \centering
    	\includegraphics[scale=0.65]{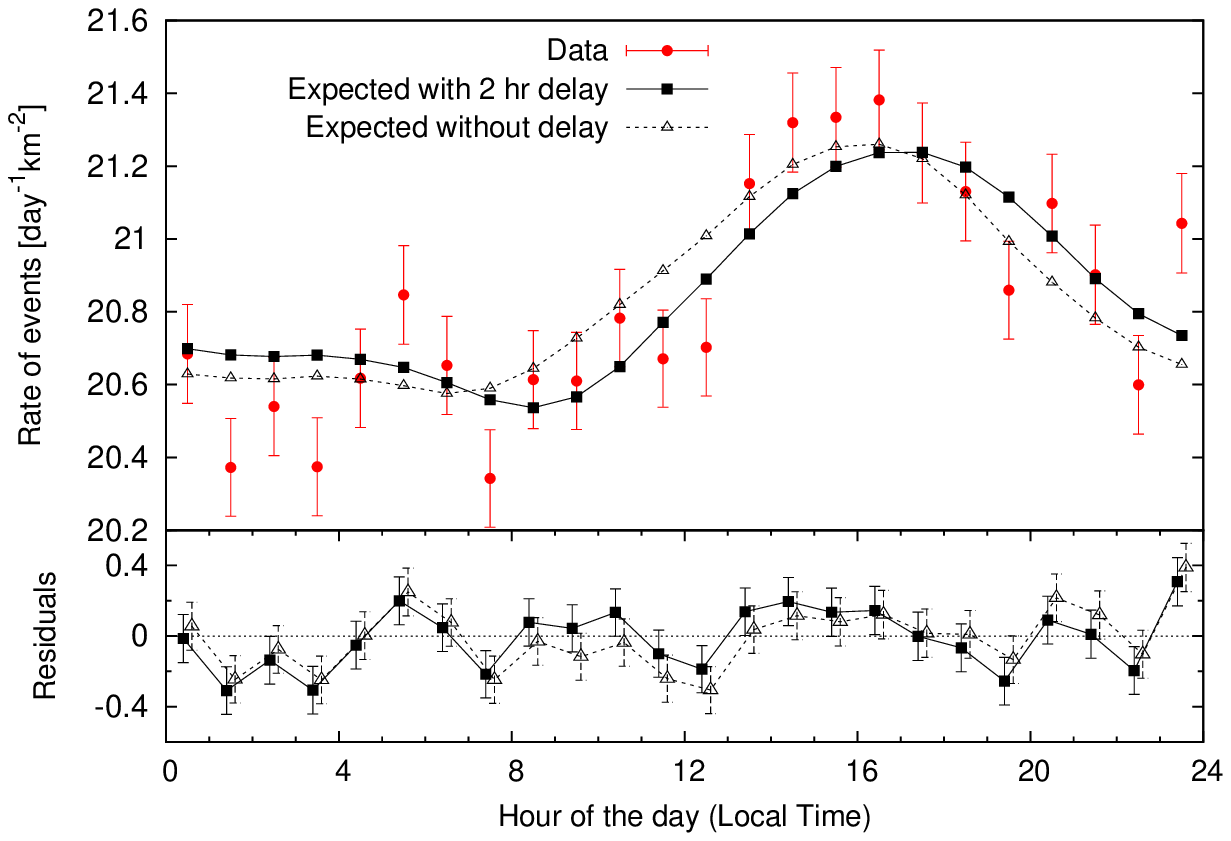}
	\end{subfigure}	
  \caption{Top panel: Rate of events per day, with $E>0.1$~EeV, for the 750~m array. The black points represent the expected rates according to the fit. Bottom panel: Average hourly measured rates and expectations from the fit (squares) and the residuals. Triangles would be the results not including the 2~h delay in the densities.}
	\label{fig:ratein}
\end{figure}

The behavior of the daily averaged rate of events (i.e., in bins of 24~h) over the time period considered is shown in Figure~\ref{fig:ratein}, top panel (red points and corresponding error bars). The rates expected from the fit are shown  as black points and the agreement with the measured rates is evident in this case too. The average hourly rate is shown in the bottom panel, with the fit including the 2~h delay (squares) leading to a $\chi^2/dof=1.76$ (while the fit without delay, shown with triangles, has  $\chi^2/dof=1.84$).

\begin{figure}[t]
  	\begin{subfigure}[t]{0.32\textwidth}
  		\includegraphics[scale=0.20]{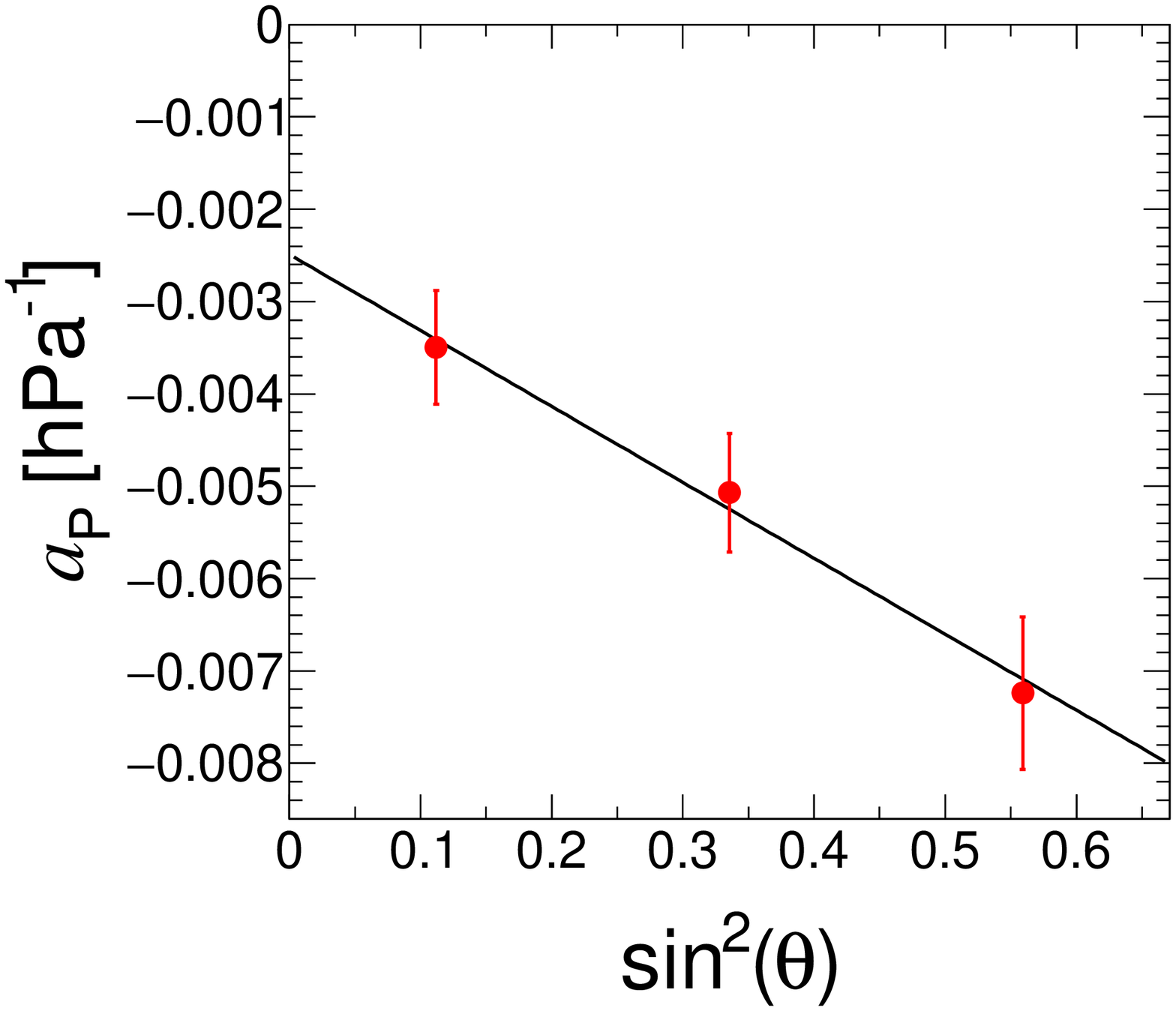}
	\end{subfigure}
	\begin{subfigure}[t]{0.32\textwidth}
    	\includegraphics[scale=0.20]{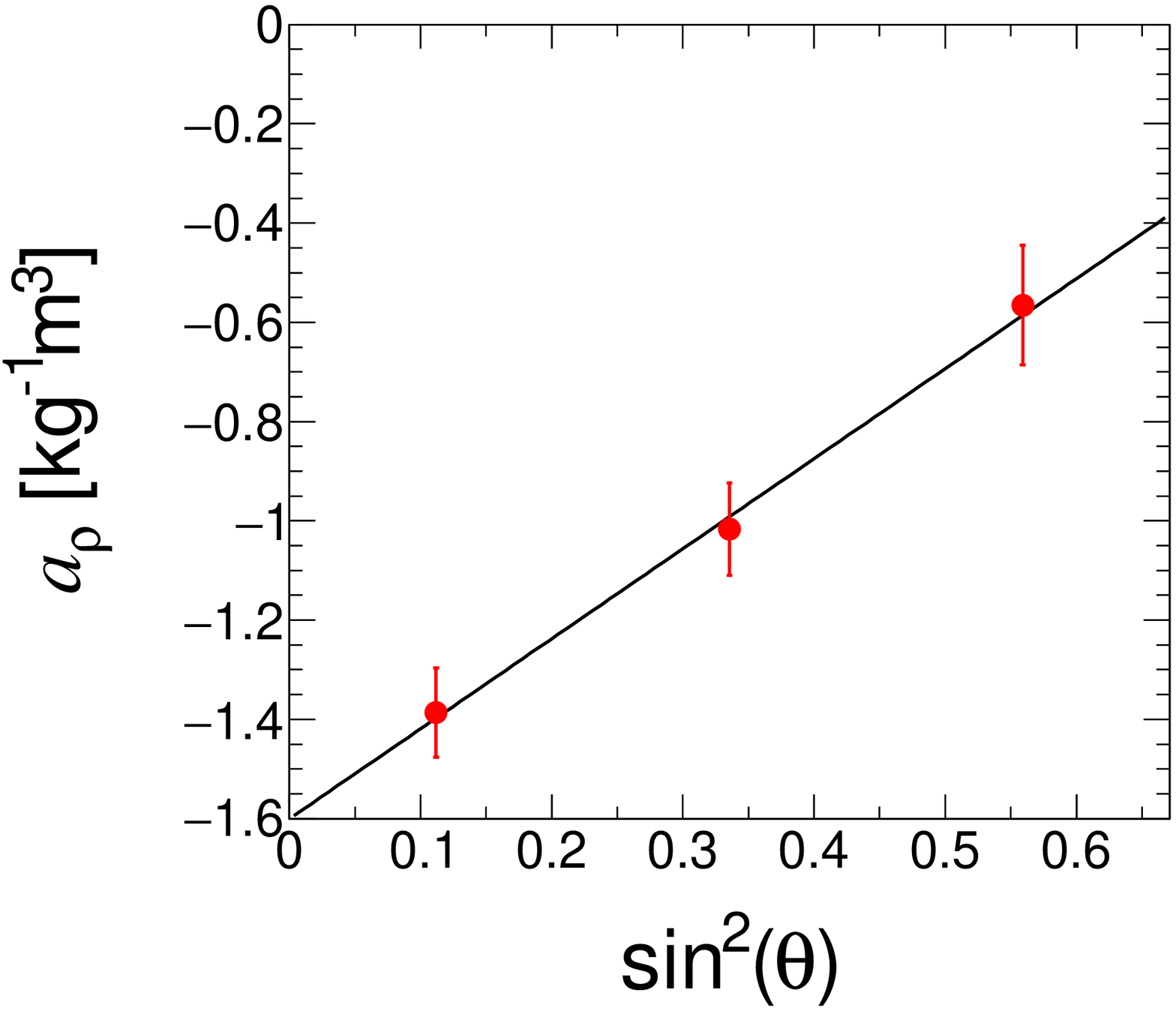}
	\end{subfigure}	
	\begin{subfigure}[t]{0.32\textwidth}
    	\includegraphics[scale=0.20]{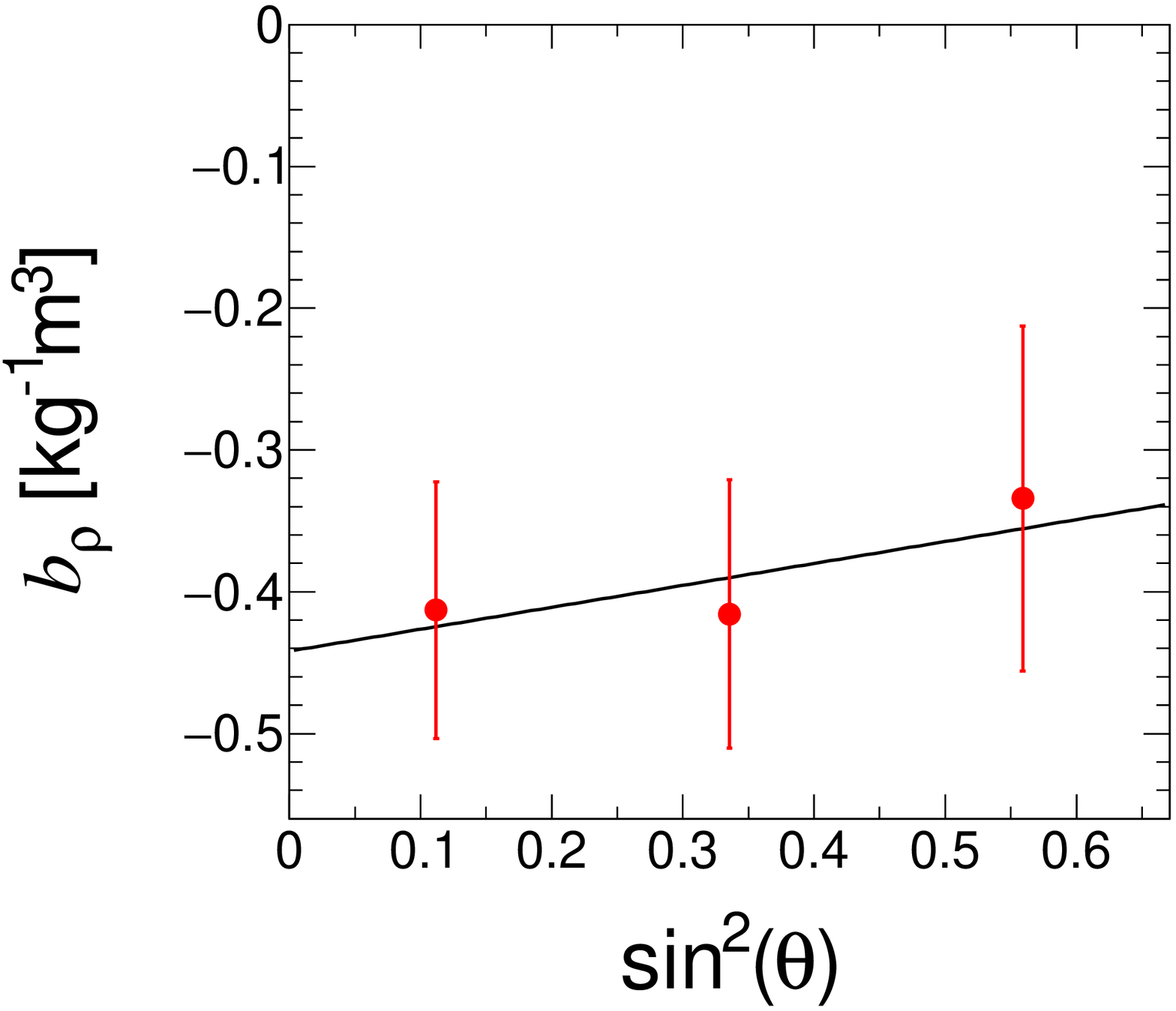}
	\end{subfigure}	
  \caption{Atmospheric coefficients $a_P$ (left), $a_\rho$ (middle) and $b_\rho$ (right) as a function of $\sin^2\theta$,   for the 750~m array. The line is the fit obtained assuming a linear dependence on sin$^2\theta$.}
	\label{fig:wcvssin2in}
\end{figure}

To analyze the dependence on zenith angle we split the 750~m array data set in equal-width bins of $\sin^2\theta$. However, due to the smaller number of events and reduced zenith angle range we use in this case just three bins. The resulting atmospheric  coefficients are thus fitted to linear functions
\begin{equation}\label{eq::fitwclin}
g(\sin^2\theta)=c_0+c_1\sin^2\theta.
\end{equation}

The results are displayed in Figure~\ref{fig:wcvssin2in}, with the best fit parameters being listed in Table~\ref{tab:wcvssinfitin}. By comparing the results with those of the 1500~m array, one finds similar features. The smaller values of $a_\rho$ found in the 750~m array case can be understood because logarithmic slope of the lateral distribution function is indeed smaller at distances from the core smaller than $\sim 700$~m than at larger distances \cite{NIM2015}. Hence, the density coefficient is smaller when the signal is evaluated at 450~m than when it is evaluated at 1000~m. Also from the comparison of the average coefficients of the 1500~m array in Eq.~(\ref{aver1500}) and those of the 750~m array in Eq.~(\ref{aver750}) we see that the relative importance of the pressure effects is larger for the 750~m array. The subdominant effect of the density variations, together with the still relatively large error bars, also explains the reduced improvement in the fit achieved with the 2~h delay.

\begin{table}[ht]
\centering
\begin{tabular}{ c c c c }
\hline\hline
 & $c_0$ & $c_1$ \\
\hline \\[-1.5ex]
$a_P~[{\rm hPa}^{-1}]$ & $(-2.5 \pm 0.8)\times 10^{-3}$ & $(-0.8 \pm 0.2)\times 10^{-2}$\\
$a_\rho~[{\rm kg}^{-1}{\rm m}^3]$ & $-1.6 \pm 0.1$ & $1.8 \pm 0.3$ \\
$b_\rho~[{\rm kg}^{-1}{\rm m}^3]$ & $-0.4 \pm 0.1$ & $0.1 \pm 0.3$ \\
\hline	\hline
\end{tabular}
\caption{Coefficients of the fit to the zenith angle dependence of the atmospheric coefficients for the 750~m array, using a linear function as in Eq.~(\ref{eq::fitwclin}).}
\label{tab:wcvssinfitin}
\end{table}

\section{Atmospheric effects on the energy reconstruction}

In the previous section we discussed how the signal $S$ at the reference distance (at 1000~m for 1500~m array and at 450~m for 750~m array) is modulated by the changes in the atmospheric variables that affect the shower development. These effects in turn imply that the rate of events above a given uncorrected signal (or energy) is modulated  by the variations in the atmospheric conditions, and we exploited this to obtain the coefficients $a_P$, $a_\rho$ and $b_\rho$ parameterizing the rate modulations. As discussed before, the coefficients modulating the signals themselves, see Eq.~(\ref{svspr}), are given by $\alpha_{P,\rho}=a_{P,\rho}/[B(\gamma-1)]$ and $\beta_{\rho}=b_{\rho}/[B(\gamma-1)]$, where $\gamma\simeq 3.29$ is the spectral index of the CR flux below the ankle energy and $B\simeq 1.02$ is related to the energy calibration. 

\begin{figure}[ht]
  \centering
  		\includegraphics[scale=0.7]{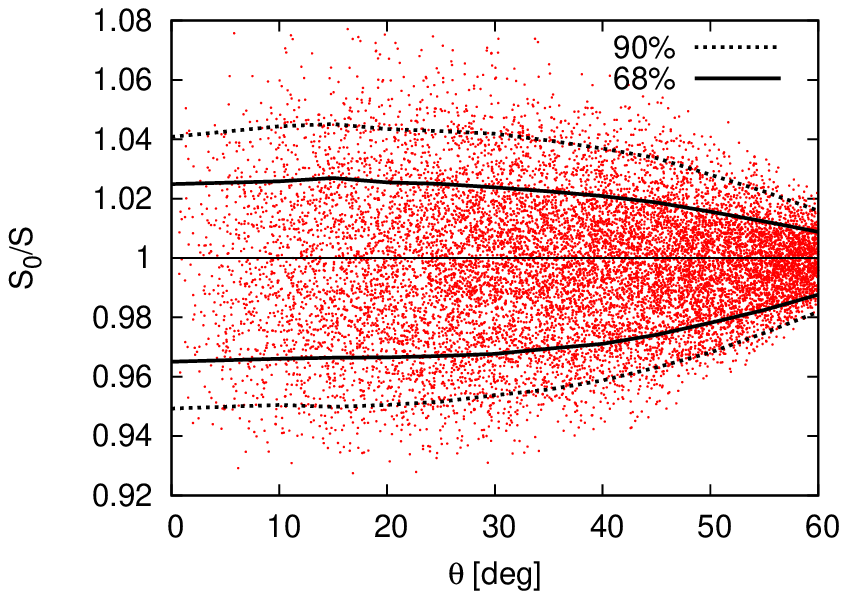}
   	\includegraphics[scale=0.7]{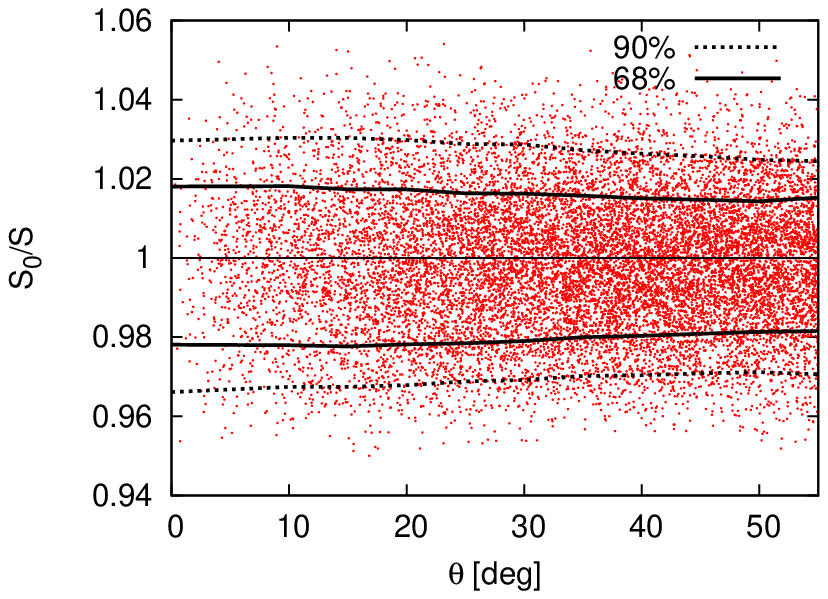}
  \caption{Ratio between the signal after ($S_0$) and before ($S$) corrections, as a function of the zenith angle, for the 1500~m array (left) and the 750~m array (right). Also shown are the contours containing 68\% and 90\% of the values at each zenith angle.}
	\label{fig:Svstheta}
\end{figure}

To illustrate the impact of the atmospheric effects, we show in Figure~\ref{fig:Svstheta} the ratio between the corrected and uncorrected signals, $S_0/S$, as a function of the zenith angle, for both the 1500~m (left panel) and 750~m (right panel) arrays. Only events recorded in 2015 are used in the two plots, as examples, with the behavior in other years being similar. Events with $E>3$~EeV are included in the left panel, and with $E>0.3$~EeV in the right panel. Given the zenith angle dependence of the atmospheric coefficients, the main effect determining the maximum amplitude of the signal correction turns out to be due to the density effect for small zenith angles, while it is instead the pressure effect for large zenith angles. Corrections to the signal, and hence to the event energy, can reach the 7\% level for extreme weather variations, but are in general at the few percent level for the bulk of the events. 

To arrive at a final expression for the energy reconstruction after correcting the signal $S$ by atmospheric effects, the next steps   would be to implement the CIC procedure in terms of $S_0$  and then to obtain the calibration coefficients for the energy. While the implementation of such procedures is beyond the scope of the present work, we discuss them here in a general way.  For definiteness we will focus below on the quantities relevant for the 1500~m array, but the discussion is completely analogous for the case of the 750~m array.

The CIC procedure assigns to each signal $S$ from a shower arriving at a given zenith angle $\theta$, a reference signal $S_{38}\equiv S/f_{\rm CIC}(\theta)$, which is the signal that would have been expected from the same CR had it arrived at a reference zenith angle of 38$^\circ$ (the median zenith of the vertical events with $\theta<60^\circ$). The function $f_{\rm CIC}$ accounts for the attenuation effects of the atmosphere, and is usually parameterized as a cubic polynomial in the variable $x\equiv \cos^2\theta-\cos^2 38^\circ$: $f_{\rm CIC}=1+bx+cx^2+dx^3$.  It is determined using only events with energies above the minimum energy for which the array is fully efficient, so as to avoid problems due to trigger issues. It is obtained by considering bins of equal width in $\sin^2\theta$ and determining the minimum values of $S$   in each bin above which there is a given number of events, the same for every bin.
 These minimum signal values are then fitted with the function $g(x)=a\,f_{\rm CIC}(x)$. In particular, the parameter $a$ turns out to be approximately the minimum value of $S$ in the central bin that includes $\theta=38^\circ$  (i.e., at $x=0$). Once atmospheric effects are accounted for, one has to consider the signals $S_0$, at the reference atmospheric conditions $P_0$ and $\rho_0$, rather than the signal $S$, and one has to repeat the CIC procedure in terms of these corrected signals. Given the linearity of the atmospheric correction it turns out that the CIC function obtained from $S_0$ is not significantly modified with respect to that obtained from $S$ if the reference atmospheric variables coincide with the overall averages. Since we here initially adopted as reference values for $P_0$ and $\rho_0$ the average ones, the CIC coefficients turn out then to be essentially unchanged. If one were to adopt instead different reference conditions,  the CIC fitting function would be expected to change because the atmospheric corrections do depend on zenith angle. 

Regarding the energy calibration, it is performed using hybrid events measured simultaneously by the surface detectors and the fluorescence telescopes. Here one relates the energy $E$ determined directly by the fluorescence telescopes, which is essentially a calorimetric measurement of the electromagnetic component of the shower, to the reference signal $S_{38}$ determined by the surface detectors.  From the observed hybrid events one fits the results to the relation $E=AS_{38}^B$  and  from it one can then assign an energy
to every EAS. Repeating this procedure in terms of the signals $S_0$ corresponding to the reference atmospheric conditions should in principle modify the calibration coefficients $A$ and $B$. The main expected effect is due to the fact that the fluorescence measurements are  performed during the night, so  that the average temperature of the hybrid events is about $6^\circ$C smaller than the average temperature of all the SD events (pressure effects should be on average quite similar during nights and days).
This implies that the average 
air density of the events used in the calibration is about 2\% higher  than the average air density for all the events (corresponding to a density difference of 0.02~kg\,m$^{-3}$). Note however that when averaging the atmospheric corrections to the energies, the one proportional to $\alpha_\rho$, involving $\rho_d-\rho_0$, will be similar in the SD and FD samples, because the daily average temperatures  are not much different in the two samples (just slightly lower for the hybrid events  due to the longer duration of the winter nights, which enhances the proportion of FD events collected during the winter). Hence,  the relevant effect when comparing the average energies will be the one due to the daily modulations determined by the coefficient $\beta_\rho$, involving $\tilde\rho-\rho_d$. Since on average $\beta_\rho\simeq -0.25$\,kg$^{-1}$m$^3$, the signals $S$ turn out to be about 0.5\% smaller on average for hybrid events than for all the events.
 These effects mostly affect the coefficient $A$ in the calibration, which should get reduced by about $\sim 0.5$\% when the calibration is expressed in terms of the signals $S_0$ at the reference atmospheric conditions. On the other hand, this is not expected to affect the coefficient $B$ significantly.
As a result, the inclusion of atmospheric effects in the calibration is expected to lower  the average of the energies assigned to the events by about 0.5\% (the precise factor depending on the zenith angle). Note that this offset is anyhow much smaller than the overall systematic uncertainties in the SD energy reconstruction, which are of about 14\%. One should also mention that changing the reference atmospheric variables $P_0$ and $\rho_0$ should affect both the CIC function and the calibration constants, besides the zenith angle dependent atmospheric coefficients. Anyhow, the final energy assignment should be essentially independent of the particular choice adopted.
A detailed implementation of these procedures will be performed in a future work.
 
\begin{figure}[t]
  \centering
  \includegraphics[scale=0.9]{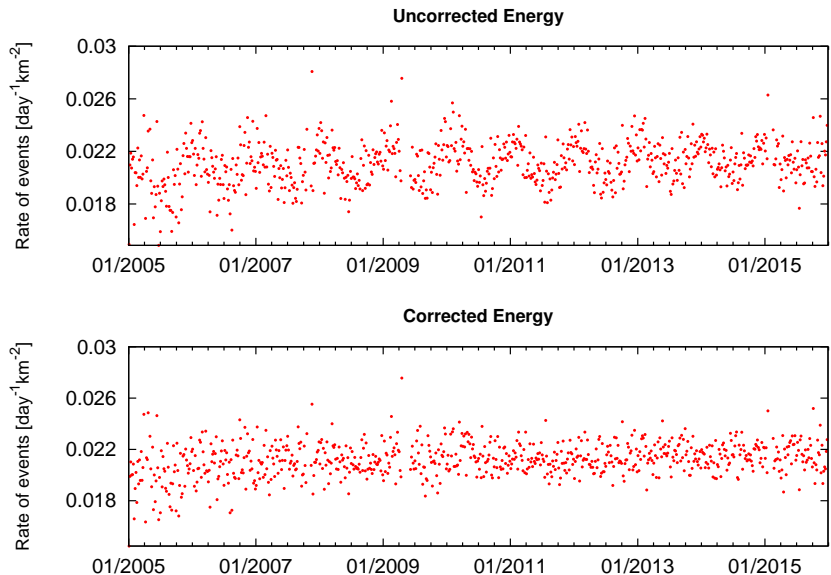}
  \caption{Daily rates before (top) and after (bottom) the correction of energies for a threshold of 2~EeV for the 1500~m array.}
	\label{fig:rateUncorCor}
\end{figure}

\begin{figure}[t]
  \centering
  \includegraphics[scale=0.9]{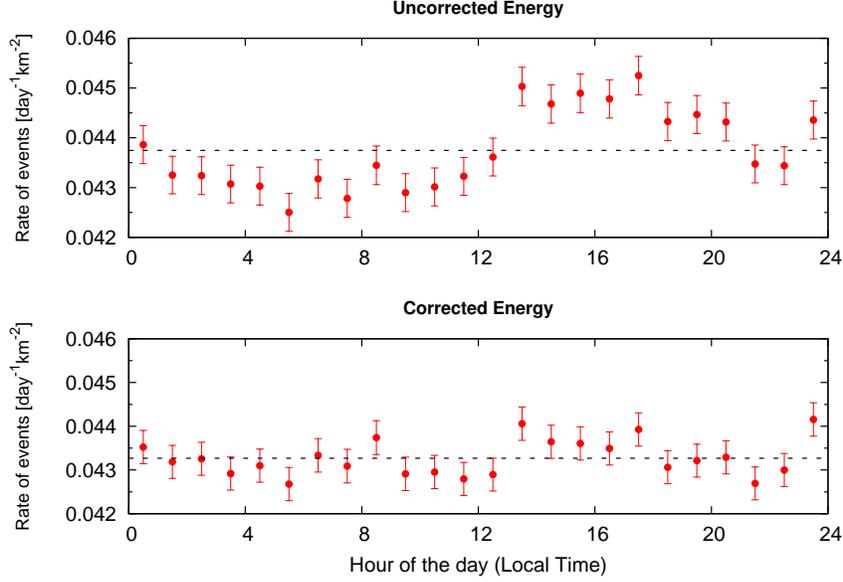}
  \caption{Hourly rates before (top) and after (bottom) the correction of energies for a threshold of 2~EeV for the 1500~m array. For guidance, horizontal lines indicate the average values.}
	\label{fig:ratehUncorCor}
\end{figure}

As a final check to verify that the atmospheric correction of the signal removes the systematic effects due to variations in the atmospheric conditions, we show in figures~\ref{fig:rateUncorCor} and \ref{fig:ratehUncorCor} the  modulation in the daily and hourly rates, respectively, above 2~EeV (which is a threshold close to full efficiency and still leading to significant number of events), using both the uncorrected energies  (top panel) and the ones corrected for  atmospheric effects\footnote{For definiteness we here  adopted $B=1.02$ and used the CIC obtained in terms of the uncorrected signals, which should provide a very good approximation.}  (bottom panel). In this last case the rates are essentially flat and
the previously existing modulation has been removed.

We have also checked that the first harmonic modulation of the rates at the solar frequency have a typical amplitude of $\sim 3.5$\% without corrections. This is due to the fact that the energies are overestimated during the afternoons, when the temperature reaches a maximum, and underestimated  during the early mornings, when the temperature drops. This artificial modulation indeed disappears when the atmospheric corrections are included in the energy assignments.

\section{Conclusions}

We have discussed in this work how to understand and correct the effects due to changes in the atmospheric variables that have an impact on the development of EAS.   We have analyzed here the data from  the two arrays of WCDs of the Pierre Auger Observatory, the 1500~m array and the 750~m array. The impact of the atmospheric effects  depends on the way in which the reconstruction is performed, and in particular on the reference distance from the shower axis adopted to obtain the signal that is used as an energy estimator, being 1000~m for the 1500~m array and  450~m for the 750~m array. We have parameterized the dependence of this signal on the  variations of air density and atmospheric pressure so as to be able to account for these effects. While the energy estimator of individual events  may be affected  at the few percent level by the variations of the atmospheric conditions, the average energy of all events is expected to be affected by no more than $\sim 0.5$\%.
The results of this work will be used in the future to improve the assignment of the energy of the  events detected with both arrays.

\section*{Acknowledgments}

\begin{sloppypar}
The successful installation, commissioning, and operation of the Pierre Auger Observatory would not have been possible without the strong commitment and effort from the technical and administrative staff in Malarg\"ue. We are very grateful to the following agencies and organizations for financial support:
\end{sloppypar}

\begin{sloppypar}
Argentina -- Comisi\'on Nacional de Energ\'\i{}a At\'omica; Agencia Nacional de Promoci\'on Cient\'\i{}fica y Tecnol\'ogica (ANPCyT); Consejo Nacional de Investigaciones Cient\'\i{}ficas y T\'ecnicas (CONICET); Gobierno de la Provincia de Mendoza; Municipalidad de Malarg\"ue; NDM Holdings and Valle Las Le\~nas; in gratitude for their continuing cooperation over land access; Australia -- the Australian Research Council; Brazil -- Conselho Nacional de Desenvolvimento Cient\'\i{}fico e Tecnol\'ogico (CNPq); Financiadora de Estudos e Projetos (FINEP); Funda\c{c}\~ao de Amparo \`a Pesquisa do Estado de Rio de Janeiro (FAPERJ); S\~ao Paulo Research Foundation (FAPESP) Grants No.\ 2010/07359-6 and No.\ 1999/05404-3; Minist\'erio de Ci\^encia e Tecnologia (MCT); Czech Republic -- Grant No.\ MSMT CR LG15014, LO1305 and LM2015038 and the Czech Science Foundation Grant No.\ 14-17501S; France -- Centre de Calcul IN2P3/CNRS; Centre National de la Recherche Scientifique (CNRS); Conseil R\'egional Ile-de-France; D\'epartement Physique Nucl\'eaire et Corpusculaire (PNC-IN2P3/CNRS); D\'epartement Sciences de l'Univers (SDU-INSU/CNRS); Institut Lagrange de Paris (ILP) Grant No.\ LABEX ANR-10-LABX-63 within the Investissements d'Avenir Programme Grant No.\ ANR-11-IDEX-0004-02; Germany -- Bundesministerium f\"ur Bildung und Forschung (BMBF); Deutsche Forschungsgemeinschaft (DFG); Finanzministerium Baden-W\"urttemberg; Helmholtz Alliance for Astroparticle Physics (HAP); Helmholtz-Gemeinschaft Deutscher Forschungszentren (HGF); Ministerium f\"ur Innovation, Wissenschaft und Forschung des Landes Nordrhein Westfalen; Ministerium f\"ur Wissenschaft, Forschung und Kunst des Landes Baden-W\"urttemberg; Italy -- Istituto Nazionale di Fisica Nucleare (INFN); Istituto Nazionale di Astrofisica (INAF); Ministero dell'Istruzione, dell'Universit\'a e della Ricerca (MIUR); Gran Sasso Center for Astroparticle Physics (CFA); CETEMPS Center of Excellence; Ministero degli Affari Esteri (MAE); Mexico -- Consejo Nacional de Ciencia y Tecnolog\'\i{}a (CONACYT) No.\ 167733; Universidad Nacional Aut\'onoma de M\'exico (UNAM); PAPIIT DGAPA-UNAM; The Netherlands -- Ministerie van Onderwijs, Cultuur en Wetenschap; Nederlandse Organisatie voor Wetenschappelijk Onderzoek (NWO); Stichting voor Fundamenteel Onderzoek der Materie (FOM); Poland -- National Centre for Research and Development, Grants No.\ ERA-NET-ASPERA/01/11 and No.\ ERA-NET-ASPERA/02/11; National Science Centre, Grants No.\ 2013/08/M/ST9/00322, No.\ 2013/08/M/ST9/00728 and No.\ HARMONIA 5 -- 2013/10/M/ST9/00062; Portugal -- Portuguese national funds and FEDER funds within Programa Operacional Factores de Competitividade through Funda\c{c}\~ao para a Ci\^encia e a Tecnologia (COMPETE); Romania -- Romanian Authority for Scientific Research ANCS; CNDI-UEFISCDI partnership projects Grants No.\ 20/2012 and No.194/2012 and PN 16 42 01 02; Slovenia -- Slovenian Research Agency; Spain -- Comunidad de Madrid; Fondo Europeo de Desarrollo Regional (FEDER) funds; Ministerio de Econom\'\i{}a y Competitividad; Xunta de Galicia; European Community 7th Framework Program Grant No.\ FP7-PEOPLE-2012-IEF-328826; United Kingdom -- Science and Technology Facilities Council; USA -- Department of Energy, Contracts No.\ DE-AC02-07CH11359, No.\ DE-FR02-04ER41300, No.\ DE-FG02-99ER41107 and No.\ DE-SC0011689; National Science Foundation, Grant No.\ 0450696; The Grainger Foundation; Vietnam -- NAFOSTED; Marie Curie-IRSES/EPLANET; European Particle Physics Latin American Network; European Union 7th Framework Program, Grant No.\ PIRSES-2009-GA-246806; and UNESCO.
\end{sloppypar}

\appendix
\section{Time delay in the temperature modulation of the atmosphere  above ground level}
\begin{figure}[ht]
  \centering
  \includegraphics[scale=0.4]{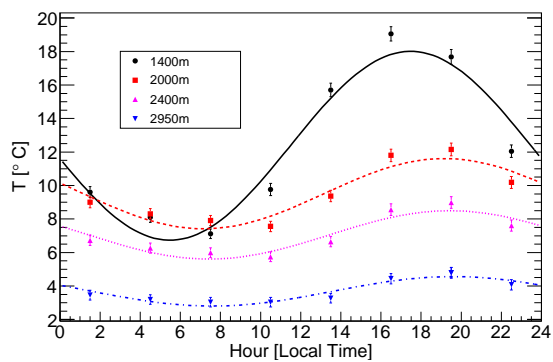}
  \caption{Temperature vs. hour of the day in Malarg\"ue, measured in Local Time, for different altitudes above sea level.}
	\label{fig:gdas}
\end{figure}

The daily temperature modulation, both its amplitude and phase, is known to depend on the considered height above ground level, and this dependence is  also different for different sites \cite{JGR05}.
It is interesting to check how the modulations in the atmospheric conditions at different heights are related to those at ground level as a function of time at the Malarg\"ue site.
For this purpose we consider data from the Global Data Assimilation System (GDAS), a global atmospheric model which provides the main state variables as a function of the height above sea level every three hours (see \cite{GDAS}). A full year from December 21 2009 to December 21 2010 of temperature data at several altitudes is taken as an example. An average of the temperatures at a given height is calculated for each available hour (shown in Figure \ref{fig:gdas} for four different altitudes, with the  1400~m one corresponding to mean altitude of the Auger site). We also perform a fit to a periodic function in time of the form $T(t)=\langle T\rangle+A\cos[\pi(t-t_d)/12{\,\rm h}]$. The results of these fits are summarized in Table~\ref{tab:Tvstfit}. It is apparent that as the height above ground level is increased the modulation  shifts to the right.
The shift between the modulation at 1400~m and that at the higher altitudes is about two hours. This is an indication that for the heights relevant for the effects on EAS observed at ground  (approximately two radiation lengths, corresponding to about $ 700\,\textrm{m}\cos\theta$ above ground) the daily modulation of the temperature gets delayed by about two hours with respect to   that measured at ground level.

One may also note from the figure how the amplitude $A$ of the daily temperature variations decreases with increasing height.

\begin{table}[ht]
\centering
\begin{tabular}{c c c c}
\hline
\hline
 height [m]& $\langle T\rangle$ [$^\circ$C] & $A$ [$^\circ$C] & $t_d$ [h]\\
\hline
1400 & $12.4 \pm 0.1$ & $5.6 \pm 0.2$ & $17.5 \pm 0.1$\\
2000 & $9.5 \pm 0.1$ & $2.1 \pm 0.2$ & $19.2 \pm 0.3$\\
2400 & $7.0 \pm 0.1$ & $1.4 \pm 0.2$ & $19.4 \pm 0.4$\\
2950 & $3.7 \pm 0.1$ & $0.9 \pm 0.1$ & $19.6 \pm 0.6$\\
\hline
\hline	
\end{tabular}
\caption{Parameters of the fit to temperature data from GDAS at different heights above sea level (see text). }
\label{tab:Tvstfit}
\end{table}

\end{document}